\documentclass[twocolumn]{aastex7}

\usepackage{amsmath}	

\usepackage{tabularx} 

\usepackage{booktabs} 

\usepackage{subcaption}

\begin{document}
\title{Long-Term Optical Follow Up of S231206cc: Multi-Model Constraints on BBH Merger Emission in AGN Disks}


 \author{P.Darc}
\affiliation{Centro Brasileiro de Pesquisas Físicas \\
 Rua Dr. Xavier Sigaud 150, CEP 22290-180, \\
 Rio de Janeiro, RJ, Brazil}
 \email{phelipedarc@cbpf.br}

\author{C. R. Bom}
\affiliation{Centro Brasileiro de Pesquisas Físicas \\
Rua Dr. Xavier Sigaud 150, CEP 22290-180, \\
Rio de Janeiro, RJ, Brazil}
\email{phelipedarc@cbpf.br}
\author{C. D. Kilpatrick}
\affiliation{Center for Interdisciplinary Exploration and Research in Astrophysics (CIERA) 
Northwestern University, Evanston, 
IL 60201, USA}
\email{phelipedarc@cbpf.br}
\author{A. Souza Santos}
\affiliation{Centro Brasileiro de Pesquisas Físicas \\
Rua Dr. Xavier Sigaud 150, CEP 22290-180, \\
Rio de Janeiro, RJ, Brazil}
\email{phelipedarc@cbpf.br}

\author{B. Fraga}
\affiliation{Centro Brasileiro de Pesquisas Físicas \\
Rua Dr. Xavier Sigaud 150, CEP 22290-180, \\
Rio de Janeiro, RJ, Brazil}
\email{phelipedarc@cbpf.br}
\author{J. C. Rodr\'iguez-Ram\'irez}
\affiliation{Centro Brasileiro de Pesquisas Físicas \\
Rua Dr. Xavier Sigaud 150, CEP 22290-180, \\
Rio de Janeiro, RJ, Brazil}
\email{phelipedarc@cbpf.br}
\author{D. A. Coulter}
\affiliation{Space Telescope Science Institute, Baltimore, MD 21218, USA}
\email{phelipedarc@cbpf.br}
\author{C. Mendes de Oliveira}
\affiliation{Universidade de S\~ao Paulo, IAG, Rua do Matão 1225, São
Paulo, SP, Brazil}
\email{phelipedarc@cbpf.br}
\author{A. Kanaan}
\affiliation{Departamento de F\'isica, \\ Universidade Federal de Santa Catarina, Florian\'opolis, SC, 88040-900, Brazil}
\email{phelipedarc@cbpf.br}
\author{T. Ribeiro}
\affiliation{Rubin Observatory Project Office, 950 N. Cherry Ave., Tucson, AZ 85719, USA}
\email{phelipedarc@cbpf.br}
\author{W. Schoenell}
\affiliation{GMTO Corporation 465 N. Halstead Street, Suite 250 Pasadena, CA
91107, USA}
\email{phelipedarc@cbpf.br}

\author{E. A. D. Lacerda}
\affiliation{Universidade de S\~ao Paulo, IAG, Rua do Matão 1225, São
Paulo, SP, Brazil}
\email{phelipedarc@cbpf.br}

\begin{abstract}

The majority of gravitational wave events detected by the LIGO/Virgo/Kagra Collaboration originate from binary black hole (BBH) mergers, for which no confirmed electromagnetic counterparts have been identified to date. However, if such mergers occur within the disk of an active galactic nucleus (AGN), they may generate observable optical flares induced by relativistic jet activity and shock-heated gas. We present results from a long-term optical follow-up of the GW event S231206cc, conducted with the T80-South telescope as part of the S-PLUS Transient Extension Program (STEP). Our search prioritized AGN-hosted environments by crossmatching the GW localization with known AGN catalogs. No candidate met the criteria for a viable optical counterpart. We explored three BBH merger scenarios predicting optical emission in AGN disks: (i) ram pressure–stripping, (ii) long term emission of an emerging jet-cocoon, and (iii) jet breakout and shock cooling. Using our observational cadence and depth, we constrained the BBH parameter space, including the remnant’s location within the AGN disk, kick velocity, and SMBH mass. Detectable flares are most likely when mergers occur at 0.01–0.1 parsecs from SMBHs of $10^7$–$10^8 M_\odot$, where short delay times and long durations best aligns with our follow-up strategy. These results provide a framework for identifying AGN-hosted BBH counterparts and guiding future multimessenger efforts.

\end{abstract}


\section{Introduction}\label{sec:introduction}

\begin{figure*}
    \centering
    \includegraphics[scale=0.2]{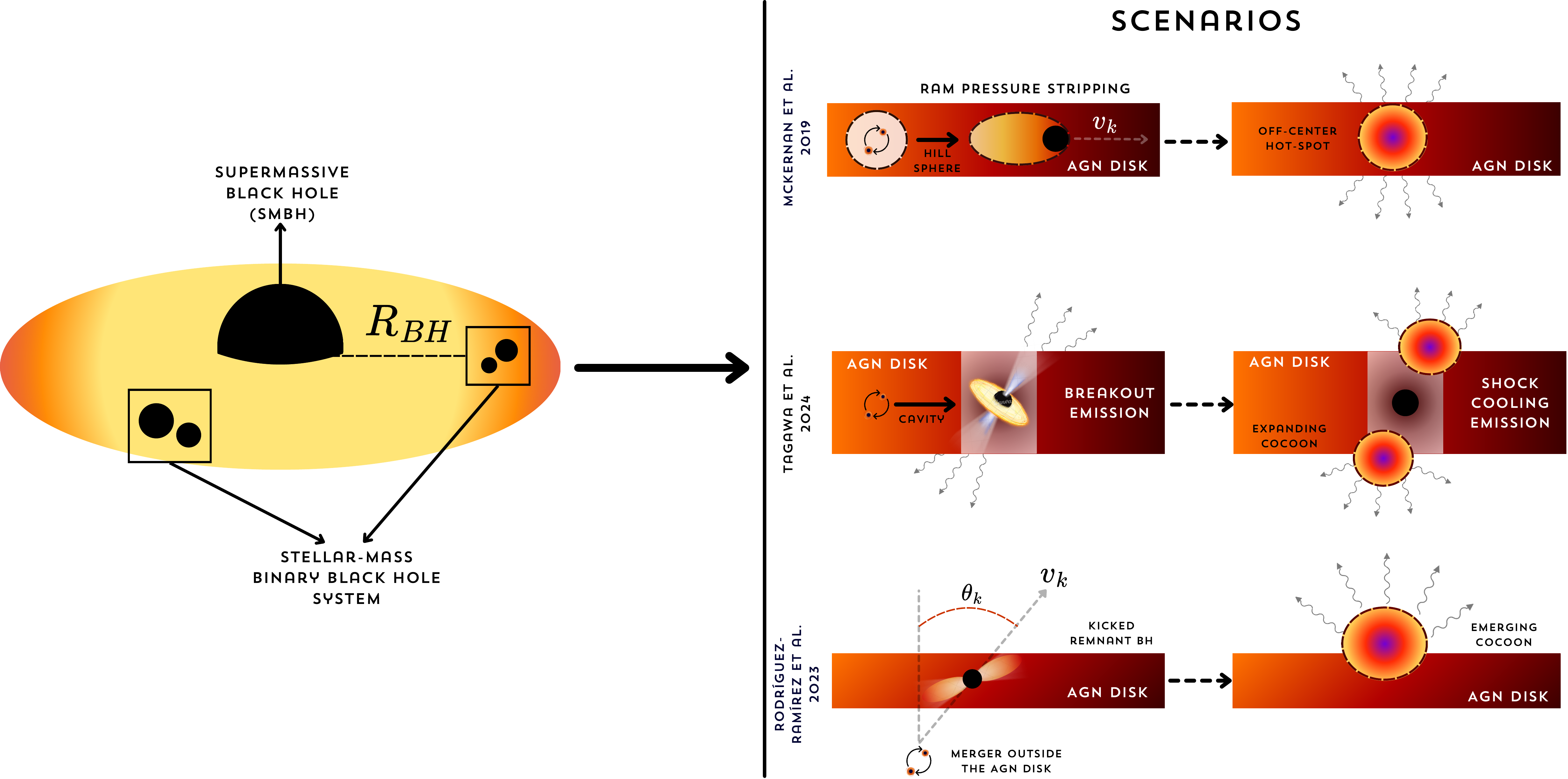}
    \caption{Multi-panel schematic showing the different 
 flare mechanisms capable of producing EM counterparts to BBH mergers in AGN disks. Each panel highlights a distinct physical scenario: top-right shows the flare model from \citet{Mckernan2019}, center-right depicts the jet breakout and shock cooling emission scenario from \citet{tagawa_phe_model}, and lower-right presents the delayed jet-driven model from \citet{juan_publicado}}.
    \label{fig:bbhschema}
\end{figure*}



The discovery of AT2017gfo, the first optical/NIR counterpart to a gravitational wave event (GW170817) \citep{Abbott2017b,Coulter2017,Soaressantos2017ApJ...848L..16S}, marked a milestone in multimessenger astronomy. This landmark produced a series of important results, including constraints on the neutron star equation of state \citep{Abbott2018}, provided an independent measurement of the Hubble constant \citep{Abbot2017c}. It demonstrated the power of combining gravitational wave (GW) and electromagnetic (EM) observations, driving ongoing efforts to identify EM counterparts to compact object mergers.

Compared to Binary Neutron Star (BNS) mergers, Binary Black Hole (BBH) mergers occur far more frequently \citep{abbott2020prospects, abbott2023}, and future GW observatories are expected to detect $10^{4-5}$ BBH mergers annually \citep{Gupta2024,Saini2024}. They can also be used to constrain $H_0$ in a method known as dark sirens \citep{bom2024dark,alfradique24}. However, while binary black hole mergers are prominent GW sources the feasibility of those sources to produce any detectable EM counterpart is object of debate, e.g. whether the merger occurs within a gas-rich environment. \citep{Bom2024, alves2024, 2025arXiv250318333L,juan_novo,juan_publicado,tagawa_phe_model,Palmese_2021bbh, 2025arXiv250502924Z, 2025arXiv250510395Z}. 


Several theoretical models propose an AGN-assisted BBH merger mechanisms that could produce observable optical, X-ray, or radio emissions \citep{Mckernan2019,Kimura2021,juan_publicado,ma2024electromagneticflaresassociatedgravitational,Chen_2024,tagawa_phe_model, juan_novo}. These models predict a broad range of emission timescales relative to the merger time, depending on the specific interaction between the BBH remnant and the AGN environment. A schematic representation of the mechanisms adopted in this work are illustrated in Figure \ref{fig:bbhschema} and further discussed in Section \ref{sec:bbh}.

Therefore, motivated by those models, recent searches have been conducted for EM counterparts to BBH events, including efforts by \cite{Ohgami_2023,Graham2023,Ahumada_2024,cabrera2024_kimuramodelgraham, Kim_2021_bbhfollowup, Kilpatrick_2021_gravcole, Doctor2019,Grado2019}. A particularly interesting case is the GW event GW190521, which showed a possible association with an optical flare  (\texttt{ZTF19abanrhr \footnote{https://lasair-ztf.lsst.ac.uk/objects/ZTF19abanrhr/}}) detected by the Zwicky Transient Facility. \cite{Graham2020}  suggested that the flare \texttt{ZTF19abanrhr}, which was observed approximately $34$ days after GW190521,  was spatially coincident with the AGN J124942.3+344929. This AGN lies within the 78\% credible region of the GW localization and is located 1.6$\sigma$ from the peak of the marginal luminosity distance distribution. The temporal and spatial coincidence, along with the observed light curve, were found to be consistent with the behavior expected from a kicked BBH merger in an accretion disk of an active galactic nucleus. This GW event produced a black hole remnant of approximately $142$ $M_{\odot}$, the heaviest black hole measured by the LIGO-Virgo-KAGRA (LVK) collaboration so far \citep{GW190521PhysRevLett.125.101102}. Notably, this event provides evidence that the heavier binary component lies within the so-called upper mass gap—roughly 50–120~$M_{\odot}$—a mass range predicted to be unpopulated due to pair-instability supernovae (PISN). The presence of such a component challenges standard stellar evolution models and suggests alternative formation pathways. These include hierarchical mergers of smaller black holes or the direct collapse of a stellar merger involving an evolved star and a main sequence companion. Such scenarios are expected to occur preferentially in dense stellar systems or AGN disks, the latter being particularly favorable not only for hierarchical mergers but also for the potential production of electromagnetic counterparts to BBH mergers.

This potential association has sparked further interest in identifying similar occurrences, as the detection of BBH EM counterparts would be of immense scientific value for both cosmology and astrophysics.  it could potentially constrain the formation environments of BBHs \citep{Veronesi2023}, serve as direct probes of AGN discs \citep{McKernan2018}, and offer a new standard-siren approach for measuring cosmological parameters \citep{Bom2024, alves2024}. However, establishing such connections is challenging due to the inherent brightness of AGN, which can obscure transient signals, and the difficulty in pinpointing the exact origin of any detected variability, especially because multiple flares may occur within the GW region around the same time, thus the definitive association remains uncertain \citep{Palmese_2021bbh,GW190521PhysRevD.108.123039}.

The term “flare” is broadly used to describe a wide array of astrophysical transients characterized by sudden brightening, encompassing phenomena such as magnetic activity in stars, solar flares, and AGN variability driven by accretion disk instabilities. However, in the current era of multi-messenger astronomy, and with growing interest in potential electromagnetic counterparts to BBH mergers occurring in AGN disk environments, we suggest an update to the terminology for this work. These transients originate from the dynamical interaction of two black holes, with a definitive emission mechanisms that remain uncertain and without any confirmed detections to date. To distinguish them from conventional AGN flares and to emphasize their unique origin, we adopt the term “Dark Flares” in this paper. This term reflects both their connection to BBH mergers and the still-unconfirmed yet theoretically motivated possibility of an electromagnetic counterpart powered by novel physical processes.



In this work, we present results from our optical follow-up of the GW event S231206cc, a BBH merger detected by the LVK collaboration on 2023-12-06 23:39:01.498 UTC. This event, with a 90\% credible localization area of 342 deg$^2$, was among the best-localized BBH mergers in the southern sky during 2023 and featured a low false alarm rate of one in 1e27 years, making it one of the most promising BBH events for electromagnetic follow-up that year. The paper is organized as follows: Section 2 reviews theoretical models for electromagnetic emission from BBH mergers. Section 3 outlines the observation strategy and provides an overview of the S231206cc event.  Section 4 describes the analysis pipeline and candidate selection and vetting process. In Section 5, we present an extension to the \texttt{Teglon} framework \citep{tegloncoulter_2021_5683508} used to model detectability and constrain BBH merger parameters using our observations. Section 6 presents our results and model-specific constraints, followed by our conclusions in Section 7. Throughout this paper, we assume a standard $\Lambda CDM$ cosmology with $H_0$ = $70 \mathrm{ km s^{-1} Mpc^{-1}}$ and $\Omega_\mathrm{m} = 0.315$

\section{Optical emission models for Binary Black hole mergers}\label{sec:bbh}

Binary black hole mergers occurring within AGN disks, where gas is present, are expected to produce electromagnetic radiation through both thermal and non-thermal mechanisms. However, only under specific conditions will these mergers generate a detectable optically bright flare. A key challenge is the intrinsic brightness of the AGN, which can outshine any potential flare associated with the BBH merger. The brighter the AGN, the more difficult it becomes to distinguish merger-associated transients from the AGN's intrinsic variability.

Recent studies have explored whether transients from BBH mergers in AGN disks could generate detectable EM counterparts. 
\cite{Mckernan2019} discussed emissions from shocks produced by collisions between gas gravitationally bound to the merger remnant and the surrounding unbound gas (gas of the AGN disk). These interactions are triggered after the newly formed black hole receives a recoil kick, induced by the anisotropic radiation of gravitational waves during the merger. Building on this scenario, \cite{Graham2020,Graham2023,cabrera2024_kimuramodelgraham} investigated the expected luminosity and delay timescales, for Bondi-accreting BH as the remnant emerges from the accretion disk due to the post-merger recoil kick.


\cite{Kimura2021} predict the existence of a pre-merger underdensity cavity around binaries embedded within AGN disks. In this scenario, the kicked merger remnant produces feedback upon re-entering the intact disk region, and the authors examined the breakout emission of the resulting outflow bubbles.
\cite{juan_publicado} and \cite{juan_novo} studied the long-term emission from disk eruptions generated by the merger remnant. In the scenario of \cite{juan_publicado}, the eruption is produced by the cocoon of a relativistic jet propagating quasi-parallel to the AGN disk. In \cite{juan_novo}, the eruption is driven instead by a transient, spherical, supercritical wind. Notably, the model of \cite{juan_novo} provides constraints on the predicted flare delay, the effective spin $\chi_{\mathrm{eff}}$, and the mass ratio $q\equiv m_2/m_1$ of the merger remnant.

\cite{tagawa_phe_model} and \cite{Chen_2024} investigate the emission produced by a jet from the BBH merger remnant embedded in an AGN disk, focusing on the emission during jet breakout, the cooling disk cocoon, and the cooling jet cocoon. The main difference between these studies is that in \cite{tagawa_phe_model},  the jet is launched immediately after the merger, while in \cite{Chen_2024}, the jet is launched only after sufficient magnetic field accumulation around the kicked remnant BH during its traversal through the AGN environment, leading to different delay times between the models.

These works suggest that the brightness, and detectability, of such flares depends mainly on two key factors: (1) the dominant emission mechanism (thermal or non-thermal) responsible for the electromagnetic counterpart, and (2) the physical properties of the remnant black hole, particularly the geometry of the merger.

We focus our analysis on BBH merger models that produce optically bright transients via thermal emission. Our primary goal is to compare theoretical light curve predictions with our follow-up observations to constrain the merger physical properties that would maximize the probability of detecting an associated flare. This section describes the models we used in this work, the physical parameters of each model, the emission mechanism and merger scenario.

\subsection{Optical/UV Flare: Ram pressure stripping of a kicked Hill sphere }

In the proposed model by \cite{Mckernan2019} (\texttt{McK19} model), the most thermally and optically luminous electromagnetic contribution following a black hole merger occurs when the remnant BH, kicked from its original position, attempts to carry along its gravitationally bound gas. As the remnant BH moves through the surrounding disk gas, it encounters an equivalent mass of gas, resulting in the loss of most of the originally bound gas due to ram pressure stripping, shock-heating the gas in its Hill sphere.  This interaction leaves behind a bright off-center hotspot with a temperature 

\begin{equation}
    T \sim \mathcal{O}(10^5\,\text{K}) \left( \frac{v_k}{10^2\,\text{km/s}} \right)^2, \quad 
\end{equation}
over a time span $O(R_{Bound}/v_k)$. The $R_{Bound}$ is the radius of gravitationally bound gas, and $v_{\kappa}$ is the kick velocity.  The dynamical time associated with the ram pressure shock, or the time it takes for the merger remnant to cross the sphere of bound gas, is given by  $t_{ram} = R_{bound}/v_{\kappa} = GM_{BH}/v_{\kappa}^3$ . Similarly to the method outlined in \cite{Graham2020,Graham2023} and \cite{cabrera2024_kimuramodelgraham}, we assume that the luminosity increase for this process scale approximately as $\sin^2(\frac{\pi}{2} \frac{t}{t_{ram}})$ until $t>t_{ram}$ , after which it decays exponentially.

Unlike previous analyses, our focus is strictly on the thermal emission from the hotspot. Therefore, we can parameterize the thermal emission by the kick velocity ($v_{k}$) and the mass of the remnant BH ($M_{BH}$), the  mass of the SMBH ($M_{SMBH}$), the accretion rate of the SMBH ($\dot{m}_{SMBH}$) and the distance of the Remnant BH from the AGN in RG ($R_{BH}$). For simplicity, in our analysis we assume a fixed accretion rate of 0.05$\times \dot{M}_{Edd}$, where $\dot{M}_{Edd}$ is the Eddington accretion rate.

\subsection{Optical Flare: emission from gas shocked due to interaction of the jets and AGN disk gas}
The \cite{juan_publicado} (hereafter \texttt{JRR-I}) and the \cite{juan_novo}'s scenario both predict a radiation signature resulting from the passage of a kicked BBH remnant through a thin AGN disk. 

The overall emission mechanism is driven by a recoiling, highly spinning BH, which forms as a result of a second or higher-generation of BBH merger occurring a few thousand Schwarzschild radii from the SMBH. After the merger, the kicked BH remnant moves into a denser region of the disk, where it begins to accrete material and launches relativistic jets that propagate quasi-parallel to the disk plane.  As these jets interact with the disk material, they create cocoons within the disk itself.  Once the cocoon emerges and expands outside the disk, thermal photons are able to escape.

\subsubsection{JRR-I: Binary Black hole merger outside the AGN disk}

Illustrated by Figure \ref{fig:bbhschema}, this scenario considers a Binary Black hole merger outside the AGN thin disk taking place at a distance $R_{BH}$ from the AGN SMBH.
The coalescence produces a kicked spinning BH remnant of mass $M_{BH}$ that enters the dense region of an AGN disk with instantaneous kick velocity $v_k$ forming an angle $\theta_k$ with respect to the disk normal.

While traveling inside the disk, it undergoes Bondi-Hoyle-Lyttleton (BHL) accretion \citep{bondihoyle} of gas from its surroundings.  This accretion is responsible by the production of relativistic jets via the Blandford-Znajek mechanism \citep{blandford-znajek}, with jet opening angle $\theta_0$ at the base. 

These jets, because of the high pressure, inflate bipolar cocoons of shocked disk material, which eventually emerges outside the disk in the form of a non-relativistic outflow. This model is focused on the thermal emission produced by photons that diffuse within the emerging cocoon (in the viewer's side) and emanates from its outer surface.

Following \cite{juan_publicado}, We adopt the standard thin disk model of \cite{shakurasunyaev_ang}, which parametrize the properties of the disk as a function of the distance to the SMBH, and its mass $M_{SMBH}$ and accretion rate $\dot{M}_{SMBH}$. Therefore, we can parameterize the theoretical merger flare light curves from a BBH merger by the mass of the supermassive black hole ($M_{SMBH}$), the mass of the kicked remnant bh ($M_{BH}$), the kick velocity ($v_k$), the kicked angle ($\theta_k$) and the distance of the remnant from the SMBH ($R_{BH}$) given in $R_g$ ($R_g = G M_{SMBH}/c^2$). The SMBH accretion rate and jet opening angle are kept fixed. Following \cite{juan_publicado}, we adopt a jet opening angle of $12$ and $\dot{M}_{SMBH} = 0.05 \times \dot{M}_{Edd}$. The detailed derivation of this model lies outside the scope of this work, further investigation regardless the thermal emission, assumptions, and parameter derivation can be found in \cite{juan_publicado}.




\subsection{Optical Flare/X-Ray: Thermal shock cooling and breakout emission }

Another promising mechanism for the production of detectable electromagnetic emission from AGN-assisted black hole mergers is the launch of a relativistic jet, similarly to the process described previously. Once the merger happens inside a cavity, the jet direction is reoriented and collides with the unshocked gas around the remnant BH. This interaction gives rise to two distinct emission mechanisms \citep{tagawa_phe_model}: breakout emission and shock cooling emission.

The \textbf{breakout emission} occurs in the early stages, shortly after the shock reaches the surface of the surrounding ejecta. As the jet propagates through the AGN disk, photons from the shocked gas begin to escape just before the jet itself breaks out of the disk. This phase of emission is characterized by a combination of non-thermal and thermal radiation, driven by the shock propagation at the jet head \citep{Chen_2024,Tagawa2023b}. 

At a later stage, when the remnant black hole can no longer sustain the jet, the system transitions into a phase dominated by \textbf{shock-cooling emission}. This emission arises on timescales of approximately $t \approx t_{\mathrm{diff}}$, which corresponds to the photon diffusion time. During this phase, the radiation originates from the adiabatic expansion of the shocked gas, leading to a cooling process as the ejecta expands. The Breakout emission is expected to be bright in X-rays and optical bands, characterized by a rapid rise and short duration. Shock cooling emission, in contrast, evolves over longer timescales, with a gradual red shift in color as the ejecta cools. 

In contrast to the methodology used by \cite{tagawa_phe_model}, which derives the model parameters based on the observed properties of the flares—such as the delay time ($t_{\mathrm{delay}}$), the duration ($t_{\mathrm{duration}}$), the observed luminosity of the flare ($L_{\mathrm{obs}}$), and the AGN luminosity—we adopt the Shakura-Sunyaev model  to parameterize the properties of the accretion disk \citep{shakurasunyaev_ang,AGNDISKMODEL_Abramowicz2013}. Specifically, we express the scale height of the AGN disk ($H_{\mathrm{AGN}}$), the density of the AGN disk ($\rho_{\mathrm{AGN}}$), and the speed of sound ($c_s$) as functions of the mass of the supermassive black hole ($M_{\mathrm{SMBH}}$), the accretion rate ($\dot{M}_{\mathrm{SMBH}}$), and the distance of the remnant black hole from the AGN in parsecs ($R_{\mathrm{BH}}$).




Therefore, the flare light curves are estimated using the luminosity evolution fo thermal emission from the shocked gas during the breakout and shock cooling emission phases, as described in Equation 25 from \cite{tagawa_phe_model} and illustrated in Figure~\ref{fig:bbhschema}. However, it is important to note that an optical excess may also arise from other non-thermal emission sources. For simplicity,  we refer to this model as the \texttt{TGW24} model. It is parameterized by five free parameters: The mass of the supermassive black hole, $M_{\mathrm{SMBH}}$;
The accretion rate of the SMBH, $\dot{M}_{\mathrm{SMBH}}$;
The mass of the remnant black hole, $m_{\mathrm{BH}}$;
The distance of the remnant BH from the AGN in parsecs, $R_{\mathrm{BH}}$;
The jet opening angle, $\theta_0$.

Following \cite{tagawa_phe_model}, we adopt the following parameter values: the opening angle of the injected jet is set to $\theta_0 = 0.2 $, the radiation efficiency to $\eta_{\mathrm{rad}} = 0.1$, the correction factor for the delay time to $f_{\mathrm{corr}} = 3$, the consumption fraction of the inflow rate to $f_{\mathrm{cons}} = 1$, and the alpha-viscous parameter to $\alpha = 0.1$. Additionally, we assume an accretion rate of $\dot{M}_{SMBH} = 0.05 \times \dot{M}_{Edd}$ for the AGN disk.

\section{Data}\label{sec:data}

\subsection{The LIGO/Virgo Event S231206cc}

The S231206cc event was reported by the LVK with a $>$ 99\% probability of being a BBH merger at 2023-12-06 23:39:01.498 UTC (MJD: 60284.98543400). After parameter estimation by \texttt{RapidPE-RIFT}\citep{riftlange2018rapidaccurateparameterinference}, the updated classification of the GW signal, in order of descending probability is BBH (100\%), NSBH ($<$1\%), Terrestrial (0\%), or BNS (0\%). The updated 50\% (90\%) confidence region spanned 90 deg2 (342 deg2) with a posterior luminosity distance of 1467 +/- 264 Mpc, or z = 0.282. S231206cc has a false alarm rate, estimated by the \texttt{gstlal} online analysis pipeline \citep{gstlalPhysRevD.95.042001}, of 1.9e-35 Hz, or about one in 1e27 years.

\subsection{T80 observations}

We interrupted the nightly schedule of the Southern Photometric Local Universe Survey (S-PLUS; \citealt{MendesdeOliveira2019}) via a Target of Opportunity (ToO) trigger for five nights.


Our criteria for triggering prioritizes events with a false alarm rate (FAR) of less than 1 per 100 years, with a strong preference for events fully-localized in the Southern Hemisphere. For BBH mergers, we impose an additional condition that the probability of the event being classified as a BBH must be equal or greater than 0.99, in order to maximize the probability of finding a merger flare.

We began our observations at 01:30 UTC on December 9, 2023, using the T80S-Cam instrument, mounted on the T80-South telescope at Cerro Tololo Inter-American Observatory. The 0.8-m robotic telescope has a 2 sq. deg field of view (FoV)  and is equipped with 12 optical bands, comprising five broad bands \textit{(g, r, i, z, u}) and seven narrow bands (J0378, J0395, J0410, J0430, J0515, J0660, J0861) from the Javalambre 12-band magnitude system \citep{MendesdeOliveira2019}. 

We imaged the 50th high-probability area of the bayestar skymap \citep{PhysRevD.93.024013} in the $i$-band with 140-sec exposures, twenty-three exposures total were taken during the first night in the $i$-band, corresponding to a 5-$\sigma$ point-source depth of $\sim$~21 mag, covering 46~deg$^2$, corresponding to $\sim$~51.1\% of the 50\% region of the updated bayestar skymap. This exposure time was chosen to maximize the probability of detecting a faint counterpart to the GW event while optimizing the sky area coverage. 

Follow-up observations were conducted on December 9, 10, 15, 16, 17 and 31, which are approximately 2, 3, 8, 9, 10 and 24 days after the GW event, respectively. The observing strategy was consistent with the first night, $i$-band with 140-sec exposures. Figure \ref{fig:skymap} illustrates our tiling over the LVK skymap, with each observation represented by a blue square. 


Unlike a kilonova — a transient powered by the radioactive decay of heavy r-process elements synthesized during a neutron star merger \citep{Metzger_2019} — which typically becomes optically faint within two weeks, requiring rapid and prompt follow-up observations, a merger flare may emerge several days to months after the gravitational wave detection. Certain models, such as those proposed by \cite{juan_novo, juan_publicado}, anticipate that a BBH merger occurring within the accretion disk of an AGN hosting a SMBH with mass $\sim 10^8 M_{\odot}$ can produce an optical merger flare peaking up to 300 days post-merger, with a prolonged emission duration exceeding 200 days. Motivated by this possibility, we conducted a final follow-up campaign between 289 and 318 days after the merger, revisiting the same fields observed during the initial search using the same observational strategy.

\subsection{Observing strategy}

For this event, the observing plan was guided by the LVK \texttt{BAYESTAR} probability map. We tiled the entire 90\% credible region and selected individual tiles based on two criteria: the number of AGNs within each tile and the observability window during the night.  To prioritize regions likely to host an electromagnetic counterpart, we utilized the Million Quasars (Milliquas \cite{milliquas2023yCat.7294....0F}) catalog, the largest existing compilation of AGNs, which includes over one million quasars, active galactic nuclei, and similar objects.

\begin{figure}[htbp]
    \centering
    \includegraphics[width=1.0\linewidth]{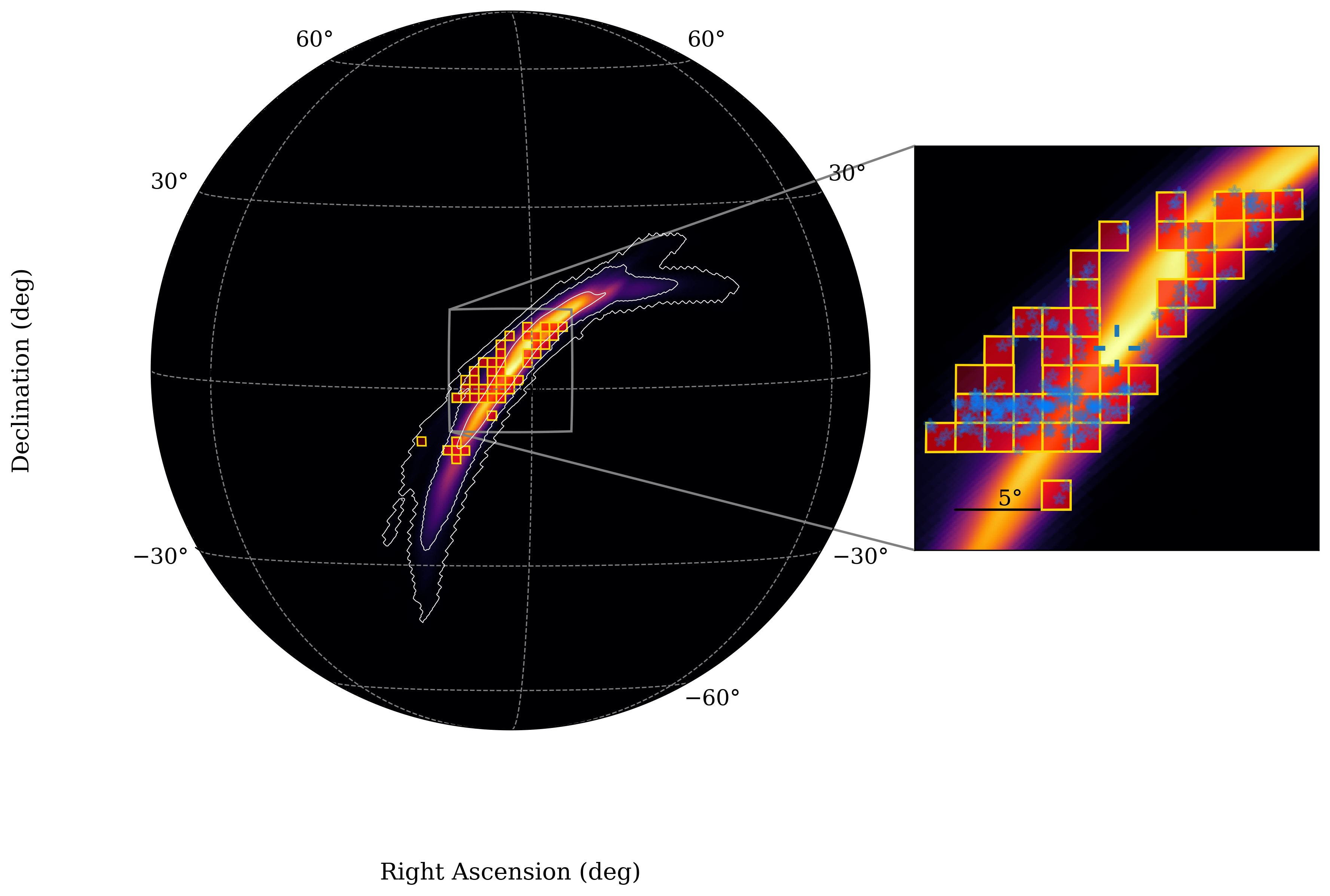}
    \caption{The LVK collaboration localization region for S231206cc. Contours correspond to the 50th (90 deg$^2$) and 90th (342 deg$^2$) percentile regions. Gold squares indicate the T80-South pointings used in our follow-up observations, while blue stars mark the positions of AGNs located within each tile. The estimated luminosity distance to the source is  $1467 \pm 264$ Mpc. }
    \label{fig:skymap}
\end{figure}

Each tile was pre-computed using \texttt{TEGLON} softaware (see Section \ref{sec:methods}) based on the \texttt{Teglon}-redistributed probability map.  Each tile's center was cross-matched with entries from the Milliquas catalog within a radius corresponding to 2 deg$^2$ (matching the T80-South field of view). We then selected AGNs with available redshift measurements and filtered those falling within a redshift range corresponding to the GW source distance, specifically, within the mean luminosity distance $\pm$ two standard deviations. Tiles were then ranked and prioritized based on the AGN density and the cumulative GW probability within the tile, ensuring maximal coverage of the relevant regions. In total, we covered 47 tiles (gold squares), encompassing 244 of the 619 AGNs (blue stars) located within the 90\% credible region of the S231206cc localization (Figure \ref{fig:skymap}).



\section{Analysis Pipeline}\label{sec:pipeline}

In this section, we describe the analysis pipeline used to search for EM counterparts to BBH mergers. We begin by outlining the S-PLUS Transient Extension Pipeline (STEP) pipeline, which handles the reduction and difference imaging processes for the search and template images. We conclude this section by detailing the candidate selection criteria employed in this study for the GW event S231206cc. 
These criteria were developed to maximize the probability of identifying a real transient associated with the GW event, followed by a final transient vetting process using forced photometry.

\subsection{S-PLUS Transient Extension Pipeline}
The STEP pipeline, which is currently operational at the Laboratório de Inteligência Artificial (LabIA) of CBPF (Brazilian Center for Physics Research), carries out transient searches in two distinct configurations:

\textbf{Nightly Scheduled Searches:} This configuration utilizes images from the on-going S-PLUS Main Survey \citep{MendesdeOliveira2019}, systematically downloading the images from the previous night and searching these fields for transient events \citep[see description in][including description of image processing and transient identification]{Santos24}. 

\textbf{Target-Of-Opportunity:} In this mode, the initial image processing is initiated immediately as the images from our search become available to enable rapid identification of transients.  STEP has contributed to targeted GW follow up in this mode for the BNS event S230615az \citep{2023GCN.33986....1S} and the BBH event S240413p \citep{2024GCN.36146....1B}.

The STEP pipeline is composed of three main stages: image reduction, difference imaging, and candidate selection. A complete description of the pipeline is provided in \citet{Santos24}. In summary, all raw images from the T80-South telescope first undergo a pre-reduction process that includes bias subtraction, flat-field correction, overscan removal, and image trimming. Once pre-processed, the images are transferred to a dedicated computing server named \texttt{Andromeda} hosted at CBPF. The \texttt{Andromeda} system is equipped with 150 TB of storage, 96 CPU cores, and 3.9 TB of RAM, supporting efficient, high-throughput data processing. All images are then processed through the reduction and difference imaging pipeline, which identifies transient sources for further analysis.

\subsubsection{STEP: Transient Candidate selection}\label{sec:step}

After completing the image processing pipeline, $\sim 19,977$ transient candidates were identified in our search. However, this initial sample includes a variety of false positives, such as solar system objects (e.g., minor planets and asteroids), foreground variable stars, and image artifacts resulting from partial PSF mismatches or minor astrometric misalignments. These artifacts are primarily due to differences between the filter systems of the science and reference images. Although astrometric misalignment is rare after calibration, it can still occur—typically caused by high proper motion stars or insufficient masking of bright sources and satellite trails in the images.

The STEP candidate selection process follows the approach of \citet{Kilpatrick_2021_gravcole}. We first crossmatched all identified transients with variable stars in the \texttt{Gaia} DR3 catalog \citep{GaiaDR32023}, rejecting matches within 1\arcsec.To minimize the number of candidates requiring visual inspection, we employed a convolutional neural network classifier based on the MobileNet architecture \citep{mobilenet}, as detailed in \citet{Santos24}, to distinguish true astrophysical transients from artifacts. After this classification stage, 13,594 candidates remained. We then applied a quality cut, retaining only sources with a signal-to-noise ratio (SNR) greater than 10, resulting in 11,959 candidates. These were further vetted via human inspection\footnote{All candidate cutouts are available through our internal user interface.}, leading to a final sample of 61 visually confirmed non-artifact sources. Lastly, we verified that no remaining candidates were associated with known solar system objects within 10\arcsec, using the Minor Planet Center database\footnote{\url{https://minorplanetcenter.net}}, and flagged known transients through crossmatching with the Transient Name Server\footnote{\url{https://www.wis-tns.org/}}. This procedure yielded the final list of 32 candidates reported in Table~\ref{tab:candidates}.

\begin{table}[h]
\centering
\begin{tabular}{@{}l r@{}}
\toprule
\textbf{Vetting Stage} & \textbf{Candidates Remaining} \\
\midrule
Initial detection               & 19,977 \\
Gaia crossmatch                & 16,685 \\
CNN filtering                  & 13,594 \\
SNR cut                        & 11,959 \\
Visual inspection              & 61 \\
Minor planet \& TNS crossmatch & 32 \\
\bottomrule
\end{tabular}
\caption{Summary of candidate filtering stages applied to the initial transient sample identified by the STEP pipeline. Each step progressively removes false positives to produce the final candidate list reported in Table~\ref{tab:candidates}.}
\label{tab:filtering_summary}
\end{table}

\subsection{S231206cc: Electromagnetic counterpart candidate selection}

For the S231206cc event, we applied a set of additional selection criteria to assist in the identification of a potential electromagnetic counterpart. These criteria were designed not only to STEP candidates but also to transients identified in other follow-up campaigns. These criteria are designed to enhance the probability of accurately finding a flare resulting from a binary black hole merger within an AGN disk environment.
\begin{enumerate}
    \item The candidate must be located inside the $90\%$ contour of the most up-to-date GW skymap.
    \item The Candidate must have its first detection, after the GW event. All the current model of optical signatures from black hole mergers expect the flare to be visible only after the merger. Therefore, we exclude any sources exhibiting variability or transient activity prior to the GW detection time. This criterion is especially important in the search for merger flares, as it helps to distinguish merger-induced emission from unrelated AGN variability.
    \item AGN matching: We expect the BBH merger to happen in an AGN disk, so we use data from the Wide-Field Infrared survey explorer \citep[WISE;][]{Wright2010wise} and Milliquas \citep{milliquas2023yCat.7294....0F} to assess whether a source could be associated to an active galactic nucleus. However, at the distance of S231206cc ($\sim 1467$ Mpc, or $z \sim 0.3$), these catalogs may miss obscured, low-luminosity, or host-dominated AGNs. While a match with a cataloged AGN is a strong indicator, the absence of a match does not exclude an AGN origin. Thus, AGN association is a valuable, but non-exclusive, criterion in identifying merger flare candidates. Additionally, Cross-matching with galaxy catalogs and checking for spatial coincidence with the center of the galaxy  provides a useful complementary indicator. 

\end{enumerate}

A key feature to discriminate candidates was to first search for a match in the Zwicky Transient Facility (ZTF) database using the FINK broker \citep{fink10.1093/mnras/staa3602}. FINK is a community broker selected to process the full stream of transient alerts from the Vera C. Rubin Observatory. Since 2020, FINK has been processing the alert stream from ZTF, allowing us to retrieve data efficiently using the right ascension (RA) and declination (DEC) coordinates within a specified radius. 

We utilized the FINK framework to search for ZTF matches within a radius of 5 arcseconds, and over a time span of 200 days before the merger and up to 300 days after the merger. This approach provided a high-speed method to identify potential candidates by cross-matching their coordinates with ZTF data. 
For candidates that did not have a ZTF match in the FINK database, we additionally made use of the ZTF forced photometry service \citep{ZTF_database_Masci2018}. This service allows us to retrieve photometric data directly from ZTF by specifying the RA and DEC of the candidates, providing valuable data for those that were not initially identified through FINK. 

However, it is important to note that ZTF does not cover the full sky accessible to T80-South. ZTF’s effective sky coverage is limited to declinations roughly between $-30^\circ$ and $+90^\circ$ due to the observatory’s location and instrumental constraints \citep{ztf2019PASP..131a8002B}. As a result, for candidates located at declinations south of $-30^\circ$, ZTF photometry, whether from FINK or the forced photometry service, is unavailable. 

In addition to FINK and ZTF service, we employed the ATLAS \citep{Atlas_Tonry2018} and ASAS-SN \citep{ASAS_Kochanek2017} forced photometry services to further analyze the candidates. ATLAS (Asteroid Terrestrial-impact Last Alert System) is a survey system designed to detect small asteroids that could potentially impact Earth. It operates in two optical bands: orange and cyan, corresponding roughly to the g and r bands of the Sloan Digital Sky Survey (SDSS). The ASAS-SN (All-Sky Automated Survey for Supernovae) is a project designed to monitor the entire sky for transient objects, such as supernovae and other variable stars. The forced photometry service from ASAS-SN enables the extraction of optical photometric measurements at precise coordinates over time, allowing us to track the apparent magnitude of transient objects.

These combined efforts, using data from ZTF and additional photometric data from ATLAS and ASAS-SN, enhance our ability to identify transient events with a higher confidence and are essential for prioritizing candidates for further spectroscopic follow-up. AGNs exhibit intrinsic variability driven by mechanisms such as disk inhomogeneities, thermal and viscous instabilities, and magnetic reconnection \citep{Graham2023}. This variability typically results in flux changes of 10–20\% relative to the baseline emission \citep{VanBerk_2004}.  flares that produce flux enhancements significantly larger than this typical variability in the background flux can plausibly be attributed to astrophysical processes beyond the AGN's intrinsic variability. Without a multi-survey photometric investigation, distinguishing between stochastic AGN variability and genuine astrophysical transients becomes challenging. A carefully curated target list, informed by the temporal evolution of each candidate across multiple surveys, reduces the risk of allocating valuable telescope time to AGNs exhibiting ordinary variability, thereby optimizing the use of limited spectroscopic resources.

Following the candidate selection process described in Section 4.1.1, from the 61 candidates, only 32 candidates were confirmed as transient (Table \ref{tab:candidates}). However, All 32 transient candidates were either known transients or were excluded based on criterion 2 described in 4.2, as they exhibited transient activity before the gravitational wave detection, within the 200-day search window. As a result, we report that none of the transients met the criteria to be considered a probable electromagnetic counterpart to the GW event.

\section{Methods}\label{sec:methods}

\subsection{Constraints on EM Counterparts to S231206cc}


Having ruled out all known optical transients as viable counterparts to S231206cc, we estimate the depth of all of our imaging and place constraints on the BBH merger properties for the different scenarios described in Section \ref{sec:bbh}.

The depth refers to the faintest magnitude of a source that can be reliably detected, which is usually defined at some significance level; in this work, we adopted a 3-sigma threshold. The limiting magnitude for a given significance level is computed using the relation:
$$
m_{\sigma} = \texttt{zpmag} - 2.5 \log_{10}(\texttt{nsigma} \times \texttt{skysig\_per\_FWHM\_Area}),
$$
which gives the magnitude of the faintest point source that can be detected with \verb|nsigma| significance, given the image’s sky noise and PSF size (seeing). The term \verb|skysig_per_FWHM_Area| approximates the total background noise within an aperture roughly matched to the FWHM of the PSF, and \verb|zpmag| is the zero point magnitude of the image.

Several recent studies have explored whether transients resulting from BBH mergers within AGN disks could account for observed optical flares and their potential association with gravitational wave events. In particular, \cite{Graham2020,Graham2023} explore whether the net luminosity and timescales for gas accretion, induced by recoil kicks, along with a thermal component from a bound gas shock \citep{Mckernan2019} and non-thermal emission from Bondi drag accretion could explain the observed optical flare \texttt{ZTF19abanrhr} and its potential association with the GW190521 event. Similarly, \cite{cabrera2024_kimuramodelgraham} applied the same methodology to the BBH S230922g gravitational wave event. 

\cite{tagawa_phe_model} used the candidates reported in \cite{Graham2023} to infer the emission properties during the early shock breakout phase and the subsequent adiabatic expansion of the shocked gas, referred to as the shock cooling emission (as previously described in Section \ref{sec:bbh} and illustrated in Figure \ref{fig:bbhschema}). However, these studies rely on information from the candidate light curves—such as time delay, rise time, and luminosity—to place constraints on the emission mechanisms.


In the following analysis, we compare between the expected in-band light curves for various BBH models and the 3$\sigma$ limiting magnitudes as a function of sky position and time relative to merger. We made use \texttt{Teglon} to calculate the likelihood that an optical counterpart to S231206cc would be detected for a given counterpart model with a set of parameters. This methodology has been adopted before by \cite{Kilpatrick_2021_gravcole} for the GW190814 event, a neutron star–black hole (NSBH) merger,  and more recently by  \cite{gravicoulter2024gravitycollectivecomprehensiveanalysis} for
GW190425, the second-ever binary neutron star merger discovered. Their focus was on emission from kilonovae, short gamma-ray burst afterglows, and linearly-evolving optical counterparts, analogous to the methodology used here for emission in a AGN disk described previously.


\texttt{Teglon} is an open-source code designed for the planning and analysis of GW events. Fundamentally, {\tt Teglon} calulates a spatially varying, 3D galaxy catalog completeness metric, and redistributes the original LVK probability to galaxies in the right 3D location for a given GW event-- resulting in a modified localization map. This function allows {\tt Teglon} to optimize EM searches by prioritizing search fields that contain high probability galaxies. 

{\tt Teglon} can also assess the detection efficiency of a given EM search by ingesting telescope pointing data (RA, DEC, 3-sigma limiting magnitude) and the cross-matching this pointing data with the corresponding, reweighted HEALPix pixel elements that each pointing contains. For each contained pixel, {\tt Teglon} retrieves the mean and standard deviation of the GW-derived, posterior distance distribution using the \verb|moments_to_parameters| function from the \textit{LIGO.skymap.distance} Python package \citep{Singer_2016}. These distance parameters are used to construct a pseudo-Gaussian distribution for each pixel, truncated at D=0 Mpc, and normalized so that the integral over the entire distribution equals the pixel's reweighted 2D probability.

To compute the detection efficiency for a given GW event, set of observations, and set of EM models (parameterized light curves in the same bands as the observations), \texttt{Teglon} re-parameterizes an observation's limiting magnitude in terms of the distance, $D_{model}$, at which we would expect to detect a modeled EM source given it's time-dependent luminosity and line-of-sight extinction,

\begin{equation}
\mu_{\text{model}, j, f} = m_{j, f} - M_{\text{model}, j, f} - A_f \\
\label{eqn:dist_mod}
\end{equation}

\begin{equation} 
D_{\text{model}, j, f} \, \text{[Mpc]} = 10^{0.2 \times (\mu_{\text{model}, j, f} - 25)}
\label{eq:dist}
\end{equation}

\noindent where in Equation \ref{eqn:dist_mod}, $\mu_{j,f}$ is the distance modulus of the $j$th pointing in the $f$th filter, and in Equation \ref{eq:dist} this distance modulus is converted to a distance.

This distance is then used in Equation \ref{eqn:net_pix_prob} as an upper limit of integration on the pixel's distance distribution, and results in a per-pixel net weight of detecting the model,

\begin{equation}
W_{\text{model}_{i,j}} = \frac{1}{\sqrt{2\pi }\sigma_{D_i}} \int_0^{D_{\text{model},j,f}} e^{-\frac{1}{2} \left( \frac{\bar{D}_i - D}{\sigma_{D_i}} \right)^2} dD,
\label{eqn:net_pix_prob}
\end{equation}

\noindent where $i$ is the pixel index, and $\bar{D}_i$ and $\sigma_{D_{i}}$ are the GW-derived distance distribution mean and standard deviation, respectively, for the $i$th pixel.

Finally, to calculate the net probability of detecting a model over all pixels and observations, in Equation \ref{eqn:net_model_prob} we take the complement of the joint probability that we do not see the source in {\it any} image,

\begin{equation}
P_{\text{model}} = \sum P_i [1 - \prod_j (1 - P_{\text{model}_{i,j}})]
\label{eqn:net_model_prob}
\end{equation}

\noindent in order to compute the probability that we detect the model in at {\it least} one image. A full description for how {\tt Teglon} computes its catalog-completeness metric, and how detection efficiencies are defined, is in Appendix A of \cite{gravicoulter2024gravitycollectivecomprehensiveanalysis}.


Therefore, we can interpret $P_{\text{model}}$ as the likelihood that we would have seen a flare with properties described by a given optical emission model for Binary Black Hole Mergers, such as MCK19, TGW24 and JRR-I. A key limitation of this method is its inability to account for non-thermal emissions from BBH mergers, as for most non thermal scenarios, there is currently no comprehensive description of the luminosity evolution over time and wavelength for such emissions. Another limitation arises from the relatively large luminosity distance of S231206cc, which reduces the effectiveness of \texttt{Teglon}’s 3D galaxy-weighted skymap redistribution. At such distances, catalog incompleteness limits the utility of galaxy-based spatial weighting.
Additionally, for every model we focus our model to be within 400 days after the merger and if the model presents a flare dimmer than the AGN base emission, this model is set to unobservable.

\section{Analysis and Discussion}

\subsection{Limits on EM Counterparts}

\subsubsection{Choice of Model Priors}


In this work, we chose priors for the simulated light curves of each BBH scenario based on physically motivated values reported in the literature.  According to \citet{vkickPhysRevLett.128.191102}, numerical relativity simulations indicate that kick velocities for precessing binary black holes (BBHs) can reach values up to $\sim 5000$ km~s$^{-1}$. 

Following the analytic fits to numerical relativity simulations provided by \citet{Lousto_2010, Lousto_2012}, the kick velocity and kick angle can be parameterized as functions of the dimensionless spin components $\bar{\chi}_1$ and $\bar{\chi}_2$, along with the masses $m_1$ and $m_2$ of the merging black holes. This parametrization enabled \citet{vkickspin10.1093/mnras/stab247}  to derive the distribution of  kick velocities for the LIGO/Virgo population in GWTC-1 \citep{Abbott_2019} and GWTC-2 \citep{Abbott_2020b}. The resulting recoil velocity distribution predominantly falls within the range of $100$–$1000$ km~s$^{-1}$ \citep{vkickPhysRevLett.128.191102, vkickspin10.1093/mnras/stab247}, which we adopted in this work to study the kicked merger remnants of a BBH with various mass ratios, spin magnitudes, and directions.

\begin{table}[h]
\centering
\renewcommand{\arraystretch}{1.5} 
\setlength{\tabcolsep}{5pt} 
\begin{tabular}{lccc}

\toprule
\textbf{Parameter} & \textbf{MCK19} & \textbf{JRR-I} & \textbf{TGW24} \\
\midrule
$v_{\mathrm{kick}}$ [km\,s$^{-1}$]       & 100–800      & 100–1000     & – \\
$m_{\mathrm{BH}}$ [$M_\odot$]             & 20–160       & 20–160       & 20–160 \\
$M_{\mathrm{SMBH}}$ [$M_\odot$]           & $10^5$–$10^9$& $10^6$–$10^9$& $10^6$–$10^9$ \\
$R_{\mathrm{BH}}$                        & $10^3$–$10^4$ $R_g$ & 300–60000 $R_g$ & $10^{-3}$–1 pc \\
$\theta_{\mathrm{kick}}$ [deg]            & –            & 1–12         & – \\
$\theta_0$ [deg]                         & –            & 12          & 12 \\
$\dot{M}_{\mathrm{SMBH}}$ [$\dot{M}_{\mathrm{Edd}}$] & 0.05         & 0.05         & 0.05 \\
\bottomrule
\end{tabular}
\caption{Parameter space ranges adopted for the three BBH optical emission models considered in this work: MCK19 \citep{Mckernan2019}, JRR-I \citep{juan_publicado}, and TGW24 \citep{tagawa_phe_model}. Each column corresponds to the adopted parameter ranges for a given model. The rows list the key physical parameters varied during the analysis: $v_{\mathrm{kick}}$ is the kick velocity of the remnant BH; $m_{\mathrm{BH}}$ is the mass of the remnant black hole; $M_{\mathrm{SMBH}}$ is the mass of the central SMBH; $R_{\mathrm{BH}}$ is the distance of the remnant BH from the SMBH, given either in gravitational radii ($R_g$) or parsecs (pc), depending on the model; $\theta_{\mathrm{kick}}$ is the angle of the kick relative to the AGN disk; $\theta_0$ is the jet opening angle (where applicable); and $\dot{M}_{\mathrm{SMBH}}$ is the SMBH accretion rate, expressed as a fraction of the Eddington accretion rate. A dash (–) indicates that the parameter is not used in that model.}
\label{tab:parameter_space}
\end{table}




The interaction between the remnant black hole and the unperturbed AGN disk also depends on the angle between the kick direction and the disk normal, $\theta_k$.  We focus our analysis on low-angle configurations, considering values from 1° to 12° \citep{juan_novo}. 


We define the prior on remnant black hole masses in our analysis based on the black hole mass distribution obtained from LIGO-Virgo-KAGRA  observations  \citep{BHMassPhysRevD.111.043020,LVKmassMandel:2018hfr}, covering a range from 20 to 160 solar masses. This assumes that the remnant black hole mass approximately follows $ M_{bh} \approx M_1 + M_2$, where $M_1$ and $M_2$ are the component masses of the binary system.

Table~\ref{tab:gwevents} presents a selection of gravitational wave events spanning a wide range of total masses that have been targeted by previous electromagnetic follow-up campaigns. The references reported are not intended to trace the origin of the mass estimates, but rather to indicate what GW events were selected for follow up. This helps illustrate that the range of mass values adopted in this work encompasses the majority of BBH events previously followed up in the search for potential electromagnetic counterparts. Some mass values may differ slightly from those reported by the Gravitational Wave Transient Catalog (GWTC). For example, while GW190521 was followed up by \citet{Graham2020}, the total mass listed here reflects the value reported by \citet{GW190521_mass2020PhRvL.125j1102A}. We also include GW190814, formally classified as a neutron star–black hole (NSBH) merger, which remains a particularly interesting case due to its extreme mass ratio—consisting of a 23.2~$M_\odot$ black hole and a 2.6~$M_\odot$ compact object.

The density within an AGN accretion disk varies with radial distance from the central SMBH. In the inner regions, typically within tens to hundreds of Schwarzschild radii ($R_{s} = 2GM/c^2$), the disk density is higher due to the stronger gravitational pull of the SMBH. As one moves outward, the overall density decreases. The precise structure of the disk depends on parameters such as the SMBH mass, accretion rate, and viscosity, as described in standard AGN disk models.
The most popular AGN disk models were presented first by \cite{shakurasunyaev_ang},  and then followed by \cite{srikogoodman10.1046/j.1365-8711.2003.06431.x} and \cite{Thompson_2005}, which assumes a geometrically thin, optically thick disks around a SMBH.  Both these models assume some heating mechanisms in the disk that marginally support the outer regions from collapsing due to self-gravity and formulate 1D sets of equations for the AGN-disk profile as a function of a number of parameters such as the mass of the central BH and the accretion rate (for a detailed reanalysis of these models see \cite{rhoagn2024MNRAS.530.3689G}). 

In our study, we aim to explore both the inner and outer regions of the AGN disk, as the observability of BBH merger-driven flares depends strongly on the local disk environment. We define the inner disk as spanning from a few hundred to a few thousand gravitational radii ($R_g$), where the ambient density is higher, while the outer disk extends to tens of thousands of $R_g$, where lower densities may allow for longer-lived, lower-luminosity emission. Accordingly, we simulate BBH mergers with $R_g$ from 300--60000, encompassing an AGN disk density of [$10^{-10}$ - $10^{-18}$] $g/cm^3$, for the JRR-I. This range allows us to probe both dense, short-timescale interactions and delayed flare scenarios at larger radii.  For the TGW24 model, this radial range is converted into parsecs to maintain consistency with its physical disk parametrization.  In the MCK19 scenario, we restrict our analysis to $R_{bh}$ to be whithin [$10^3$ - $10^4$] $R_g$, and the recoil kick to 100--800 km~s$^{-1}$, consistent with the energy required for observable thermal emission.


Several works, including \cite{Greene_2007,Bentz_2015,Kozłowski_2017},  have shown that the masses of  SMBHs in AGN predominantly range from $10^6$ — $10^9$ M$_{\odot}$. Based on these estimates, we constrain our analysis to this mass range across all models.  
The chosen flat priors for each parameter and their corresponding models are summarized in Table~\ref{tab:candidates}. These priors were selected to be as uniform and agnostic as possible to avoid biasing the parameter space.

\subsubsection{Factors Affecting Detectability of each Model}

As the total area covered by our observations comprises $\sim 26\%$ of the total two-dimensional probability in the latest S231206cc map, the strongest constraints on EM counterparts are at this significance level. The detection probabilities for BBH models from the JRR-I, TGW24, and MCK19 depends on various model-specific merger parameters. Consequently, for a given black hole mass ($M_{\mathrm{BH}}$) and kick velocity ($v_{\mathrm{kick}}$), the detection probability is not a fixed value but rather a distribution influenced by variations in other parameters.

To visualize this, we present the detection probability efficiency plot using two approaches. First, for each combination of $M_{\mathrm{BH}}$ and $v_{\mathrm{kick}}$ (or any two parameters $\theta_1, \theta_2$), we compute the maximum probability across all possible parameter configurations. This approach highlights the best-case detection scenario for each pair of $M_{\mathrm{BH}}$ and $v_{\mathrm{kick}}$, allowing us to gain insight in the full parameter space. In this plot, dark regions indicate that no observable flare is expected for any possible model for a given $M_{\mathrm{BH}}$ and $v_{\mathrm{kick}}$, while bright regions signify that at least one model within that parameter space yields a high detection probability.

In the second approach, we compute the mean detection probability across all parameter configurations for each combination of $M_{\mathrm{BH}}$ and $v_{\mathrm{kick}}$. This method allows us to identify trends in the detection probability across the parameter space, revealing regions where the probability of detecting an observable flare increases systematically. By examining these trends, we gain insight into the parameter regimes where observational signatures of BBH mergers are more likely to be detected given the models we consider.

There are four main reasons why certain regions in the efficiency plot exhibit low detection probabilities:

\begin{enumerate}
    \item \textbf{Instrument Sensitivity Limits:} The flare is too dim to be detected given the observational capabilities of our observations. 
    
    \item \textbf{Observing Strategy Constraints:} The flare is bright enough to be detected but falls outside the observational window of our follow-up strategy. 
    
    \item \textbf{AGN Background Obscuration:} The flare’s brightness is overpowered by the AGN’s baseline emission, making it undetectable by our analysis pipeline. 
    
    \item \textbf{Disk Thickness Constraints:} The AGN disc must be sufficiently thick at the location where the black hole remnant plunges in. If the disc is too thin —as described by Equation 4 in \cite{juan_publicado}— the remnant does not have enough time to accumulate sufficient accreting material, preventing jet formation. This condition is restricted to the JRR-I model.
\end{enumerate}


\subsection{The \textbf{\citeauthor{Mckernan2019} (MCK19)} Model}

\begin{figure}
    \centering
    \includegraphics[width=1\linewidth]{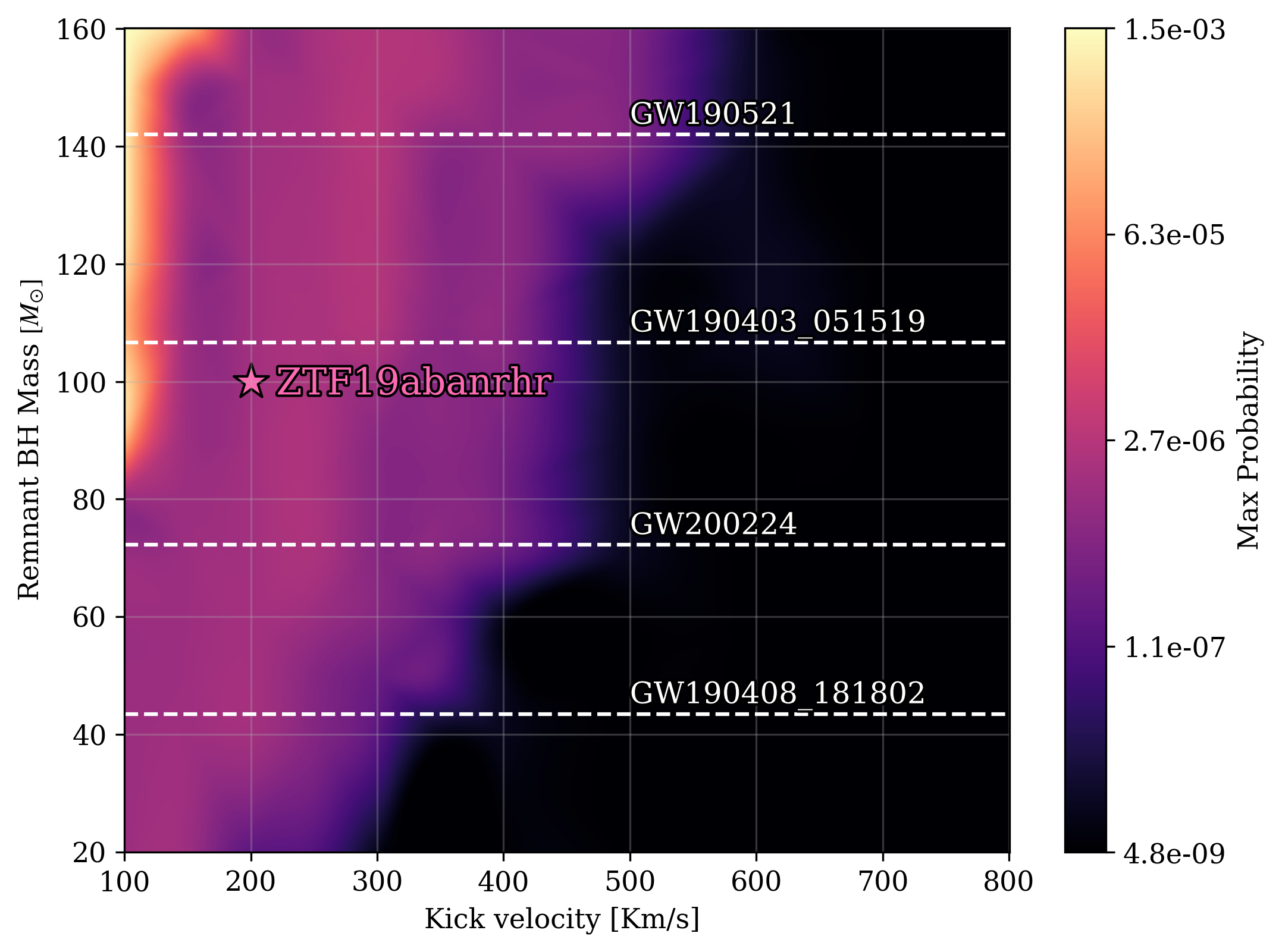}
    \caption{ Maximum detection probabilities for BBH models from MCK19 model as a function of remnant black hole mass ($M_{\mathrm{BH}}$) and kick velocity ($v_{\mathrm{kick}}$). The color scale is logarithmic, with a maximum probability of 0.15\%. The GW events depicted in the plot are detailed in Table 2, encompassing events from previous follow-up studies \citep{Graham2020,Graham2023,Kim_2021_bbhfollowup,Ohgami_2023}. The star symbol denotes the flare candidate ZTF19abanrhr, reported by \citet{Graham2020}, for which the parameters $v_{\mathrm{kick}} = 200$ km~s$^{-1}$ and $M_{\mathrm{BH}} = 100\,M_{\odot}$ were assumed.}
    \label{fig:mck19}
\end{figure}

Figure \ref{fig:mck19} presents the maximum detection probability for the MCK19 model. As expected from previous analysis from \cite{Graham2020,Mckernan2019}, the highest probability region ($p_{max} \sim 0.15\%$) corresponds to kick velocities $\sim  100-200 $ km~s$^{-1}$, which are associated with a hotspot temperature of approximately 100,000--200,000~K, peaking in the UV bands. 

Our analysis suggests a correlation between remnant BH mass and recoil velocity, indicating that for more massive black holes, our observations are sensitive to higher recoil velocities (Figure \ref{fig:mck19}). This is closely linked to the flare delay time, given by $t_{\mathrm{delay}} = G M_{\mathrm{BH}} / v_{k}^3$ \citep{Graham2020}. For example, a remnant black hole of $160 M_\odot$, receiving a kick $ v_{k} \gtrsim 500$ km~s$^{-1}$ would produce a delay time of less than three days. Such short delays could cause the resulting optical flare to peak before the first scheduled follow-up observation, significantly limiting its detectability within typical survey cadences.  our analysis shows that late-time flares are more readily detectable. For the same black hole mass, a lower kick velocity ($v_{k} \lesssim 100$ km~s$^{-1}$) result in flare emission that peaks around 300 days post-merger, often accompanied by higher luminosity, aligning well with the timescale of our late-time observations. Roughly, all flares in the MCK19 model suffers from instrumental sensitivity limitations and observational strategy constraints. Specifically, mergers with remnant masses below \(80\,M_\odot\), regardless of the kick velocity, tend to produce optical counterparts with apparent magnitudes \(\gtrsim 23\), placing them near or below the detection threshold of most current wide-field optical surveys.

While prior works (e.g., \cite{Graham2020}) adopted a kick velocity of 200 km~s$^{-1}$ based on flare timing arguments, our analysis derives observational constraints on the viable kick velocity range by comparing model brightness to the detection limits of our survey. It is important to note that our estimates are based solely on thermal emission from the post-merger hotspot, a relatively low-luminosity component, which results in a maximum detection probability of less than 0.15\%.

\begin{figure*}
    \centering
    \begin{subfigure}{0.4\textwidth}
        \includegraphics[width=1\linewidth]{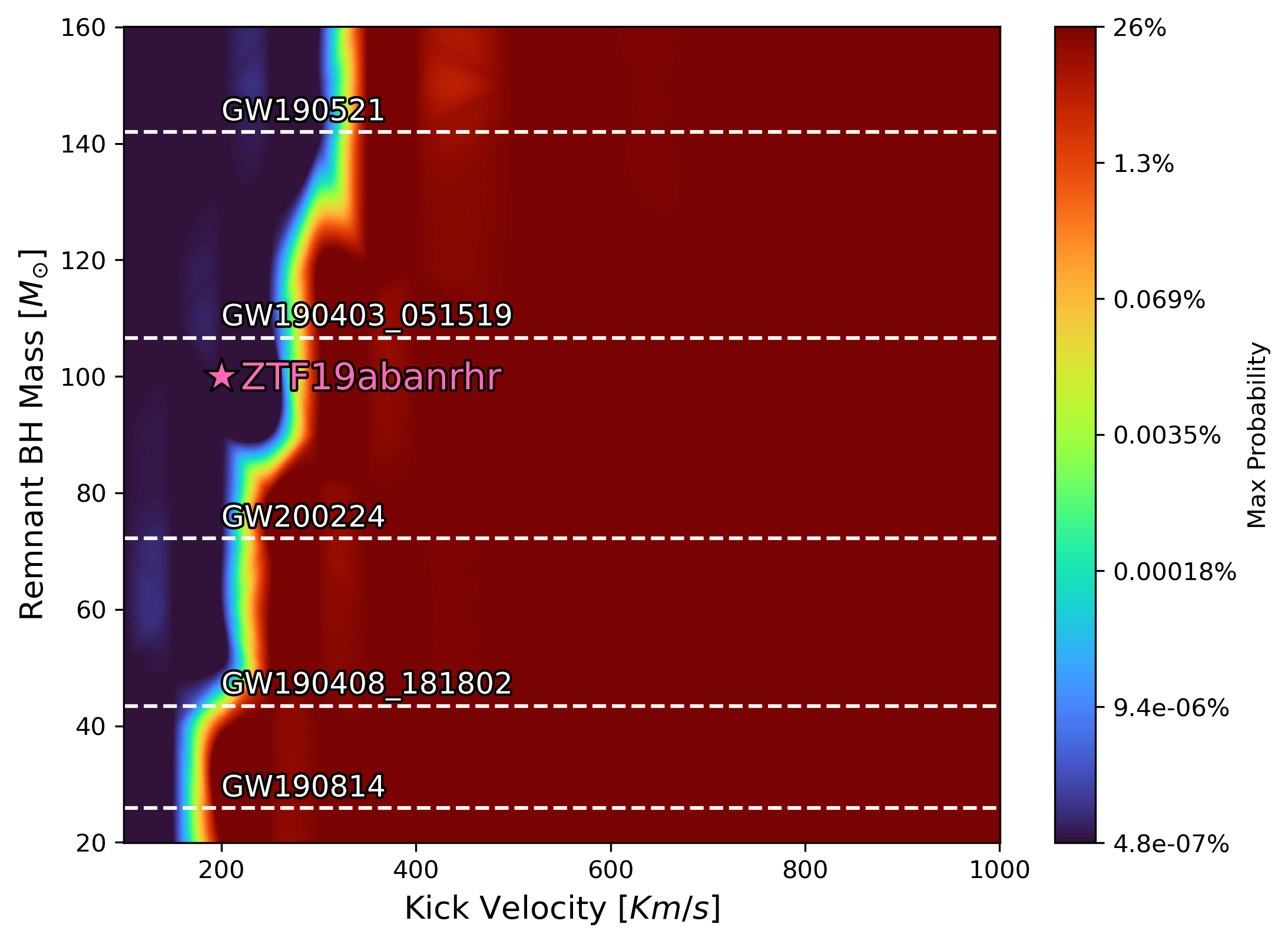}
        \caption{}
        \label{fig:jrr1_vka}
    \end{subfigure}
    \begin{subfigure}{0.4\textwidth}
        \includegraphics[width=1\linewidth]{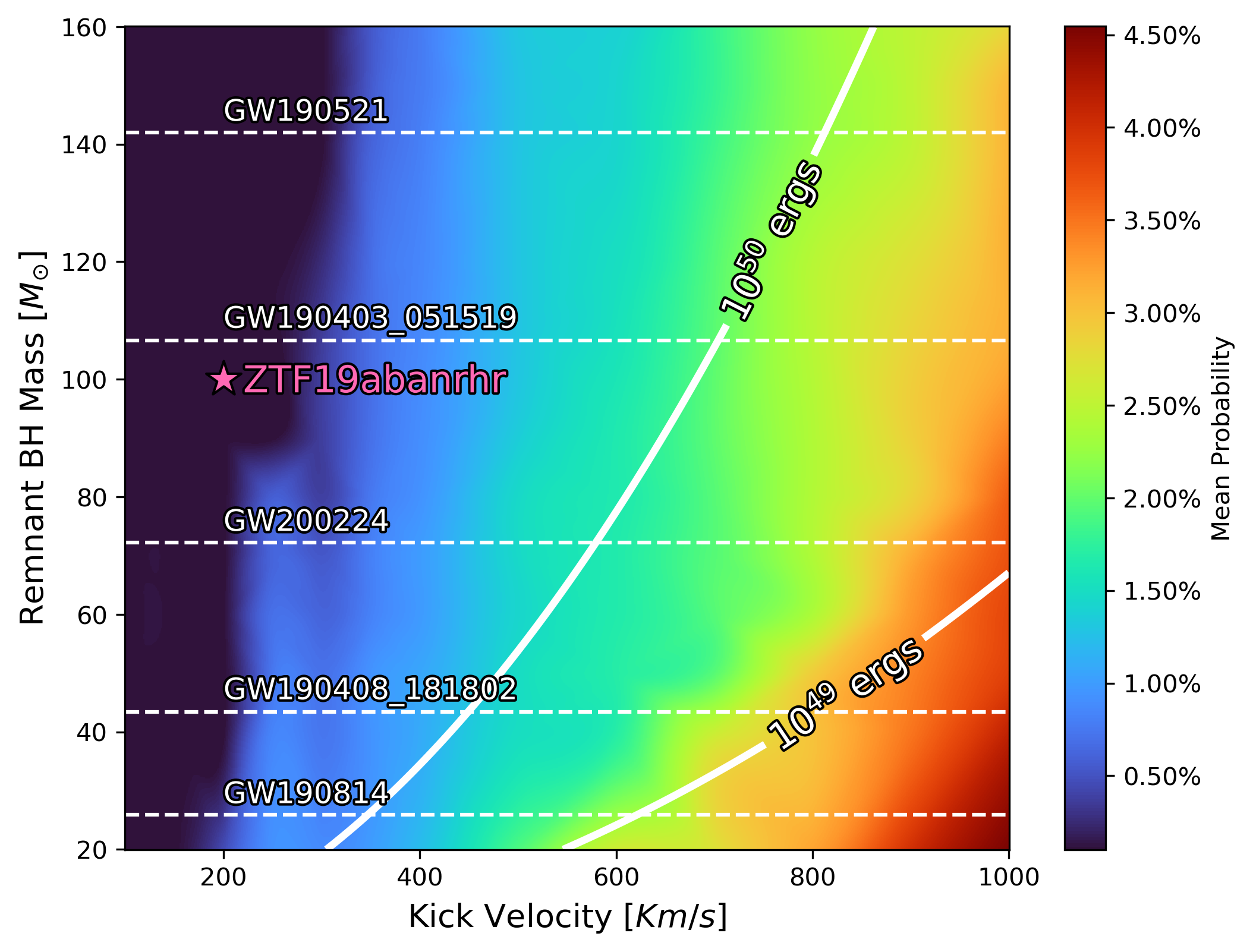}
        \caption{}
        \label{fig:jrr1_vkb}
    \end{subfigure}
    \caption{Detection probabilities for BBH models from JRR-I model as a function of black hole mass ($M_{\mathrm{BH}}$) and kick velocity ($v_{\mathrm{kick}}$). (Left) Maximum detection probability across all parameter configurations, illustrating the best-case detection scenario for each $(M_{\mathrm{BH}}, v_{\mathrm{kick}})$ pair.  Dark blue regions denote areas where no observable flare is anticipated ($P \sim 10^{-5}\%$), while bright red regions indicate at least one model with a high detection probability.
    (Right) Mean detection probability across all parameter configurations, revealing  trends and regions with increased likelihood of detecting an observable flare. Dark blue regions represent areas with low mean probability ($P < 1.0\%$), whereas red regions correspond to higher mean probabilities, suggesting a greater likelihood of observation.
    The GW events depicted in the plot are detailed in Table \ref{tab:gwevents}, encompassing events from previous follow-up studies \citep{Graham2020,Graham2023,Kim_2021_bbhfollowup,Ohgami_2023}. The star symbol denotes the flare candidate ZTF19abanrhr, reported by \citet{Graham2020}, for which the parameters $v_{\mathrm{kick}} = 200$ km~s$^{-1}$ and $M_{\mathrm{BH}} = 100\,M_{\odot}$ were assumed. The energy contours are computed assuming a SMBH mass of $5 \times 10^7 M_\odot$, a kick angle of $\theta_k = 3^\circ$, and a merger located at a distance of $6000 R_g$ from the SMBH.
    }
    \label{fig:jrr1_vk}
\end{figure*}

\begin{figure*}
    \centering
    \begin{subfigure}{0.4\textwidth}
        \includegraphics[width=\linewidth]{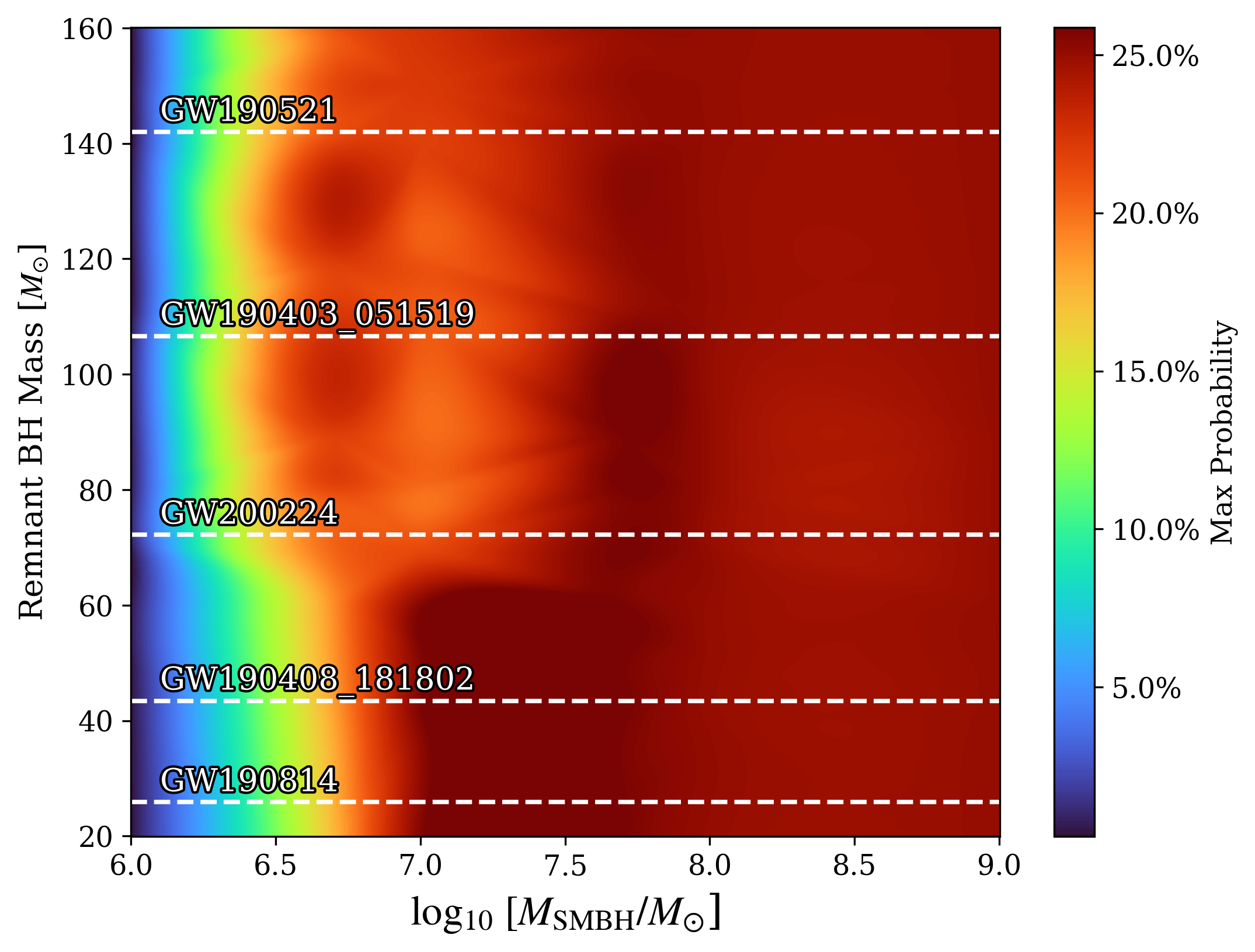}
        \caption{}
        \label{fig:jrr1_massa}
    \end{subfigure}
    \begin{subfigure}{0.4\textwidth}
        \includegraphics[width=\linewidth]{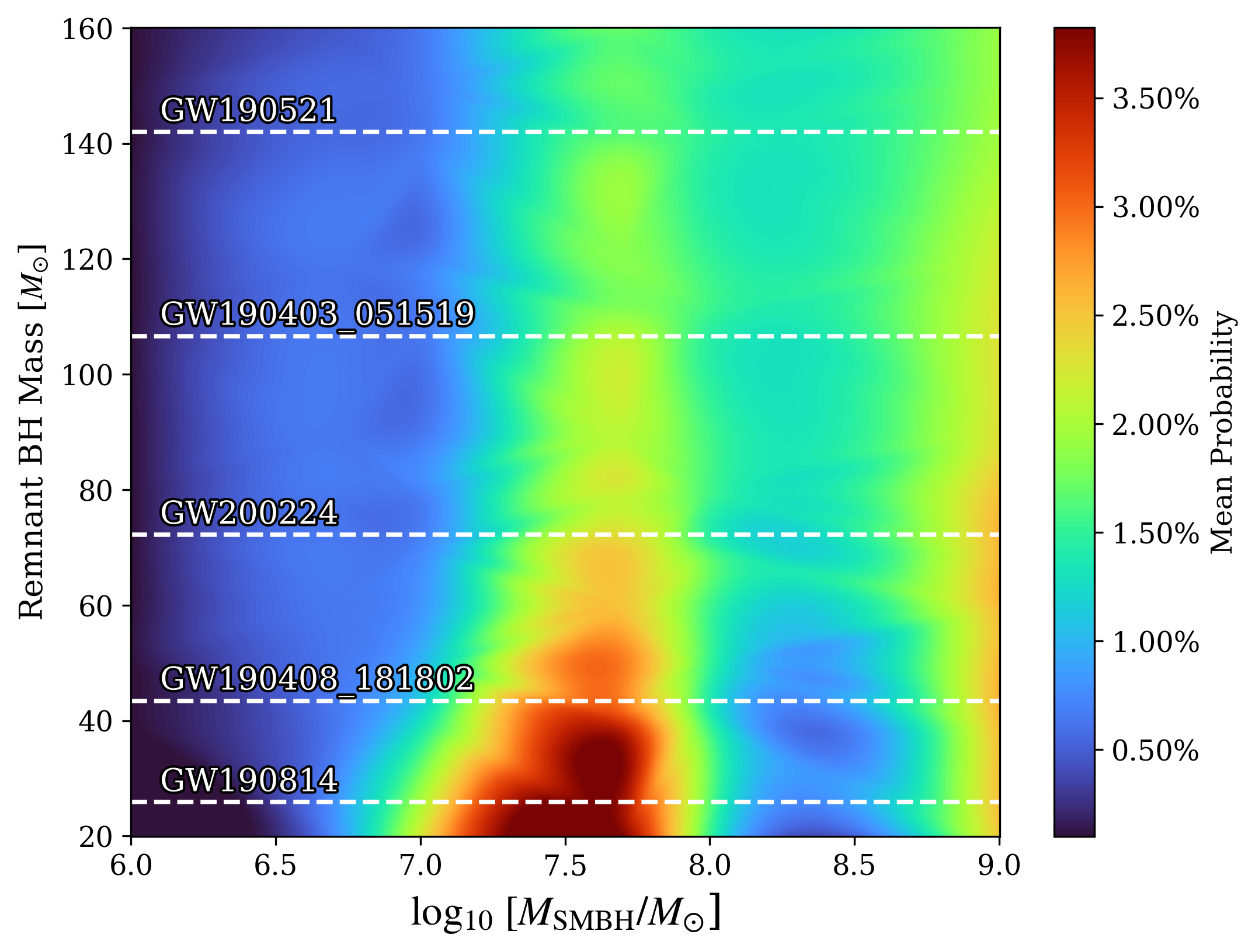}
        \caption{}
        \label{fig:jrr1_massb}
        
    \end{subfigure}    
    \caption{Detection probabilities for BBH models from JRR-I model as a function of remnant black hole mass ($M_{\mathrm{BH}}$) and log of the SMBH mass ($\log_{10}(M_{\mathrm{smbh}}/M_{\odot})$). 
    (Left) Maximum detection probability across all parameter configurations, illustrating the best-case detection scenario for each $(M_{\mathrm{BH}},\log_{10}(M_{\mathrm{smbh}}/M_{\odot}))$ pair.  Dark blue regions denote areas with a low probability of an observable flare ($P \sim 5\%$), while red regions indicate at least one model with a high detection probability.
    (Right) Mean detection probability across all parameter configurations. Dark blue regions represent areas with low mean probability ($P < 0.50\%$), whereas red regions correspond to higher mean probabilities, suggesting a greater likelihood of observation.    
    }
    \label{fig:jrr1_mass}
    
\end{figure*}

\begin{figure}
    \centering
    \includegraphics[width=1.0\linewidth]{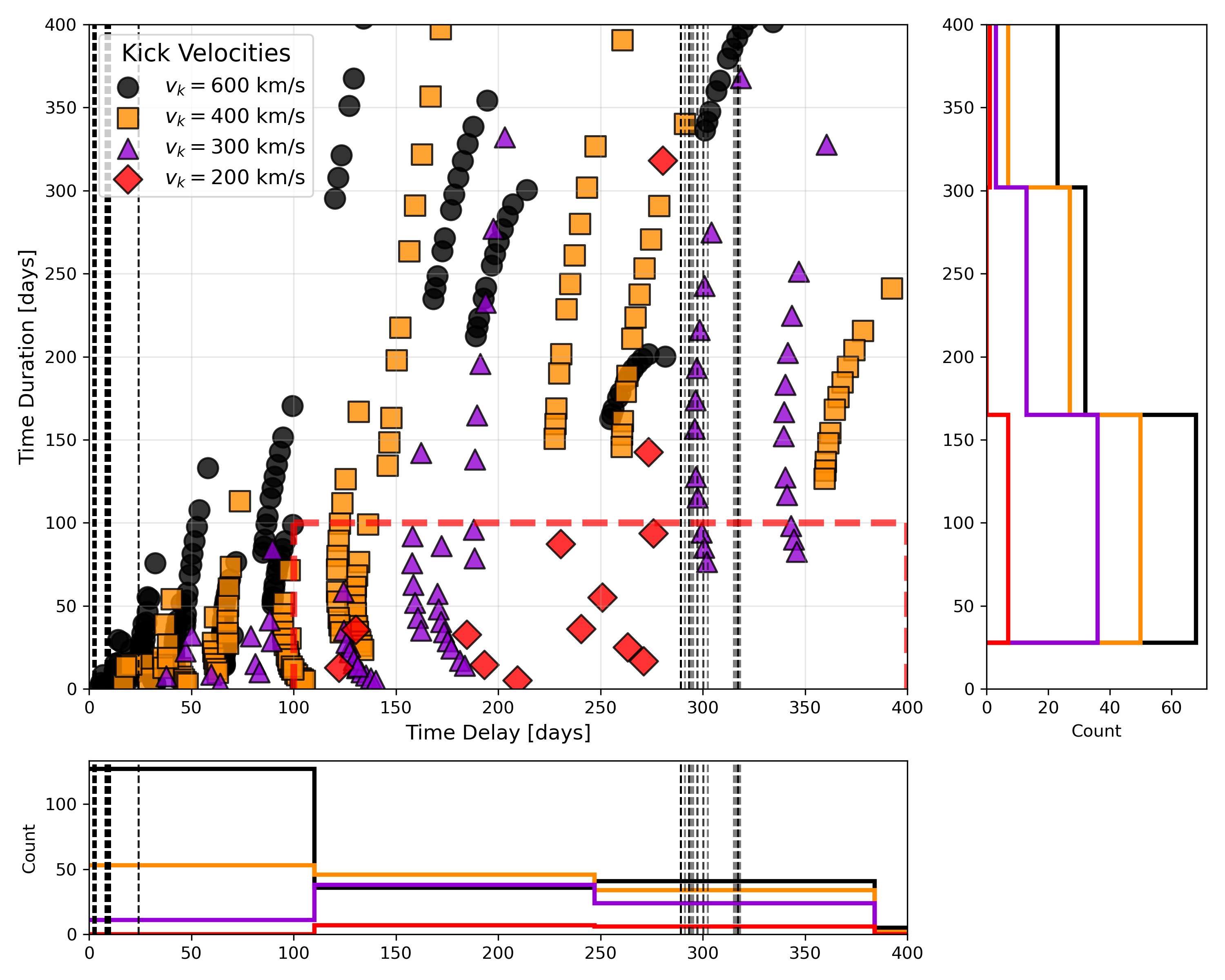}
    \caption{Scatter plot and histograms of the time duration and delay for different kick velocities ($v_k$) in km~s$^{-1}$, with a fixed $\theta_k = 0.3$. Each color represents a different kick velocity: black ($v_k = 600$ km~s$^{-1}$), orange ($v_k = 400$ km~s$^{-1}$), purple ($v_k = 300$ km~s$^{-1}$), and red ($v_k = 200$ km~s$^{-1}$). Higher kick velocity values were omitted for visual clarity. The red dashed box encloses models with a time delay greater than 100 days and a duration of less than 100 days, while the grey dashed vertical line indicates the observations made by T80. The side histograms display the distributions of delay times and durations for each kick velocity.}
    \label{fig:jrrIdelay}
\end{figure}

\subsection{The 
\textbf{\citeauthor{juan_publicado} (JRR-I) } Model
}

The disk thickness constraint and the observing strategy are the main reasons why flares are not detected when kick velocities are too low (\( v_{\mathrm{kick}}  < 200 \) km~s$^{-1}$) as shown by Figure~\ref{fig:jrr1_vk}.  Even though low kick velocities produce higher-energy flares ($E_{flare} > 10^{50}$ ergs) the time required for jet formation becomes too long (see Equation~2 from \cite{juan_publicado}). As a result, the disk has to be sufficiently thick or the remnant does not spend sufficient time within the disk to form an efficient jet, preventing the production of detectable cocoons. Even when the jet is succefully produced, it fall outside our observability window. We can visualize this effect by plotting the time-delay distribution for different kick velocities (\( v_{\mathrm{kick}} \)) in models that successfully produce a flare. From Figure~\ref{fig:jrrIdelay}, we observe that for mergers assuming \( \theta_k = 0.3 \), those with a kick velocity of \( v_{\mathrm{kick}} = 200 \) km~s$^{-1}$ produce flares only after approximately 100 days. In contrast, higher kick velocities (\( v_{\mathrm{kick}} > 400 \) km~s$^{-1}$) result in flares that require rapid follow-up (\( t_{\mathrm{delay}} < 50 \) days) and exhibit extended durations (\( t_{\mathrm{duration}} > 100 \) days).  


As expected, the kick velocity significantly influences the flare's starting time (delay time), amplitude, and duration. Generally, higher kick velocities correspond to longer flare durations and shorter delays, thereby enhancing the probability of detection in our observations, as shown in Figure~\ref{fig:jrr1_vka}. This trending is further illustrated in Figure \ref{fig:jrr1_vkb}, which demonstrates that our observational strategy is most sensitive to mergers with higher kick velocities and lower remnant masses.

We now shift our focus to another key observable parameter: the mass of the supermassive black hole in the active galactic nucleus. Given its role in shaping the disk environment and the propagation of outflows, the SMBH mass is expected to impact both the flare production condition and its observational detectability.

Figure \ref{fig:jrr1_mass} illustrates this effect by comparing detection probabilities across varying SMBH masses. As shown in Figure \ref{fig:jrr1_massa}, AGNs with $M_{SMBH} \sim  10^{6}$ are the least likely environments for producing detectable flares ($p_{max} < 5\%$). This is primarily due to two factors: (1) Disk geometry and jet formation constraints, which from all the models produced in that region, only about 20\% of models in this SMBH mass regime successfully produce flares brighter than an apparent magnitude of 24, largely due to inefficient shock interaction in thinner disks. (2) Short flare durations, most events in AGNs with \(M_{\mathrm{SMBH}} \sim 10^6\,M_\odot\) last less than 25 days, reducing the likelihood of detection within our observation strategy.

Figure \ref{fig:jrr1_massa} also shows that for AGNs with $ 5 \times 10^{6} \gtrsim   M_{SMBH} \lesssim 10^{9}$, at least one model achieves a detection probability greater than 20\%, reinforcing the idea that higher-mass AGN hosts are more favorable environments for detecting electromagnetic counterparts to BBH mergers.

    
\begin{table}
    \centering
    \begin{tabular}{|l|l|l|} \hline 
         \textbf{GW ID}&  \textbf{Total Mass} & \textbf{Source} \\ \hline 
         GW190521&  142$\substack{+28\\-16}$ & \cite{GW190521_mass2020PhRvL.125j1102A}\\ \hline 
         GW200224& 72.2$\substack{+7.2\\-5.1}$ & \cite{Ohgami_2023}\\ \hline 
         GW190403\_051519&  106.6$\substack{+26.7\\-23.6}$& \cite{Graham2023}\\ \hline 
         GW190814&  25.9$\pm$1.3 & \cite{Morgan2020}\\ \hline 
 GW190408\_181802& 43.4$\substack{+4.2\\-3.0}$ & \cite{Kim_2021_bbhfollowup}\\ \hline
    \end{tabular}
    \caption{List of GW events that were targeted by previous electromagnetic follow-up campaigns, along with their total masses and corresponding references.}
    \label{tab:gwevents}
\end{table}

Figure \ref{fig:jrr1_massb}  reveals a significantly enhanced mean probability — by an order of magnitude — of detecting mergers in AGNs with $M_{SMBH} \sim 10^{7.5}$ when the black hole remnant mass is less than $40 M_{\odot}$. This suggests that, on average, models within this mass range and host produce flares that are ten times more likely to be observed compared to other regions of parameter space. This enhancement arises due to a selection effect in the flare formation criteria: all failed flares are set as unobservable, leading to variations in the number of contributing models across different host masses. 
The $M_{\rm SMBH} \sim  5 \times 10^{7}~M_{\odot}$ region maximizes the number of parameter configurations that successfully generate bright observable flares (apparent mag $\leq $ 24.0), making it the most favorable for detection.

In conclusion, considering the JRR-I model, our detectability analysis suggest that targeting AGNs with SMBH masses in the range of $10^{7-8} M_\odot$ maximizes the likelihood of flare detection. In contrast, for lower-mass SMBHs ($\sim 10^6 M_\odot$), rapid follow-up observations are essential to optimize detection efficiency due to the shorter flare duration times. Meanwhile, mergers occurring in AGNs with $ 10^9$~$M_{\odot}$ exclusively produce flares with delay times exceeding 300 days post-merger, as a result, these events are only detectable within our current extensive observation strategy.


    





\begin{figure*}[h]
    \centering
    \begin{subfigure}{0.4\textwidth}
        \includegraphics[width=\linewidth]{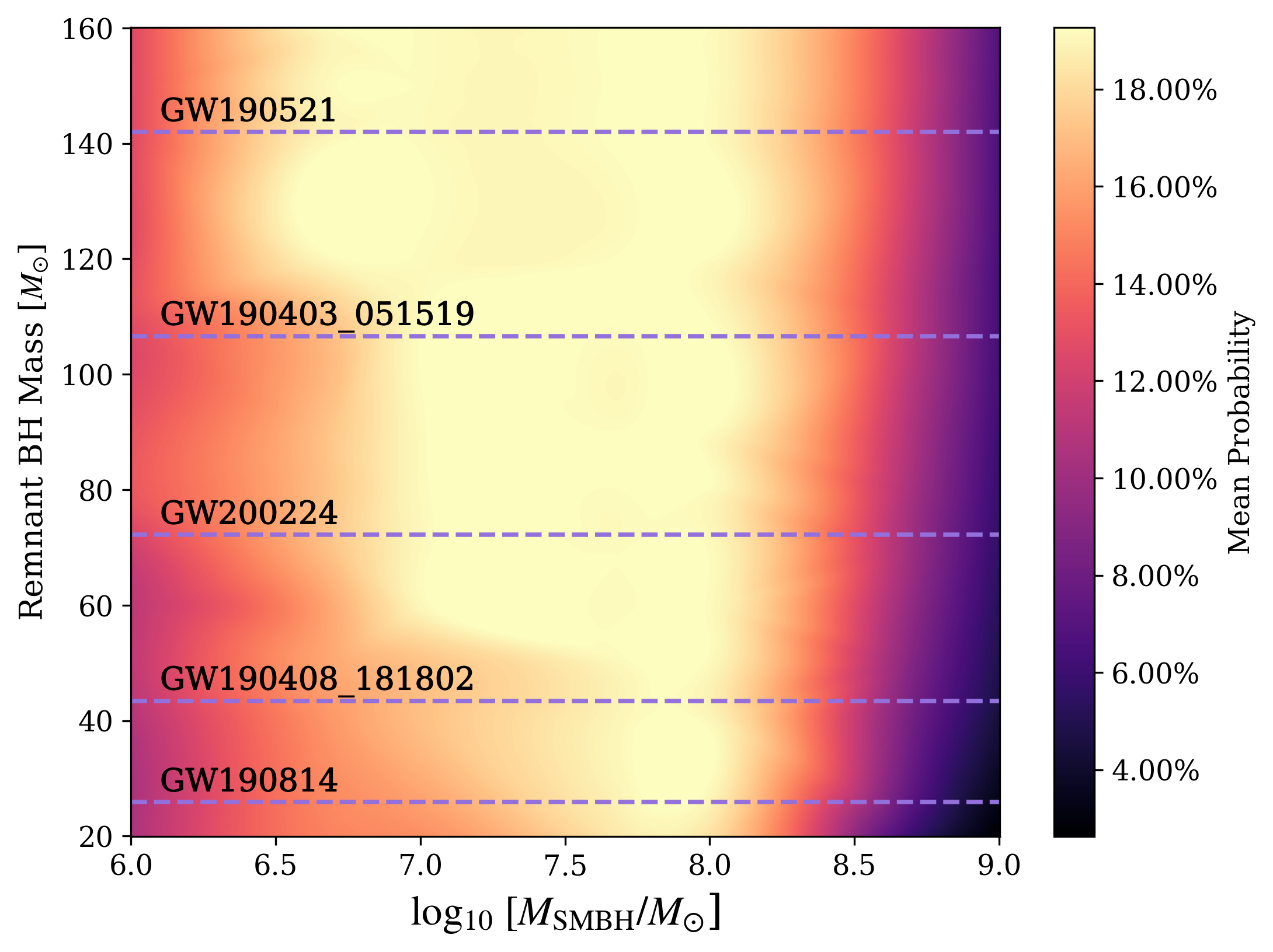}
        \caption{}
        \label{fig:tgw24a}
    \end{subfigure}
    \begin{subfigure}{0.4\textwidth}
        \includegraphics[width=\linewidth]{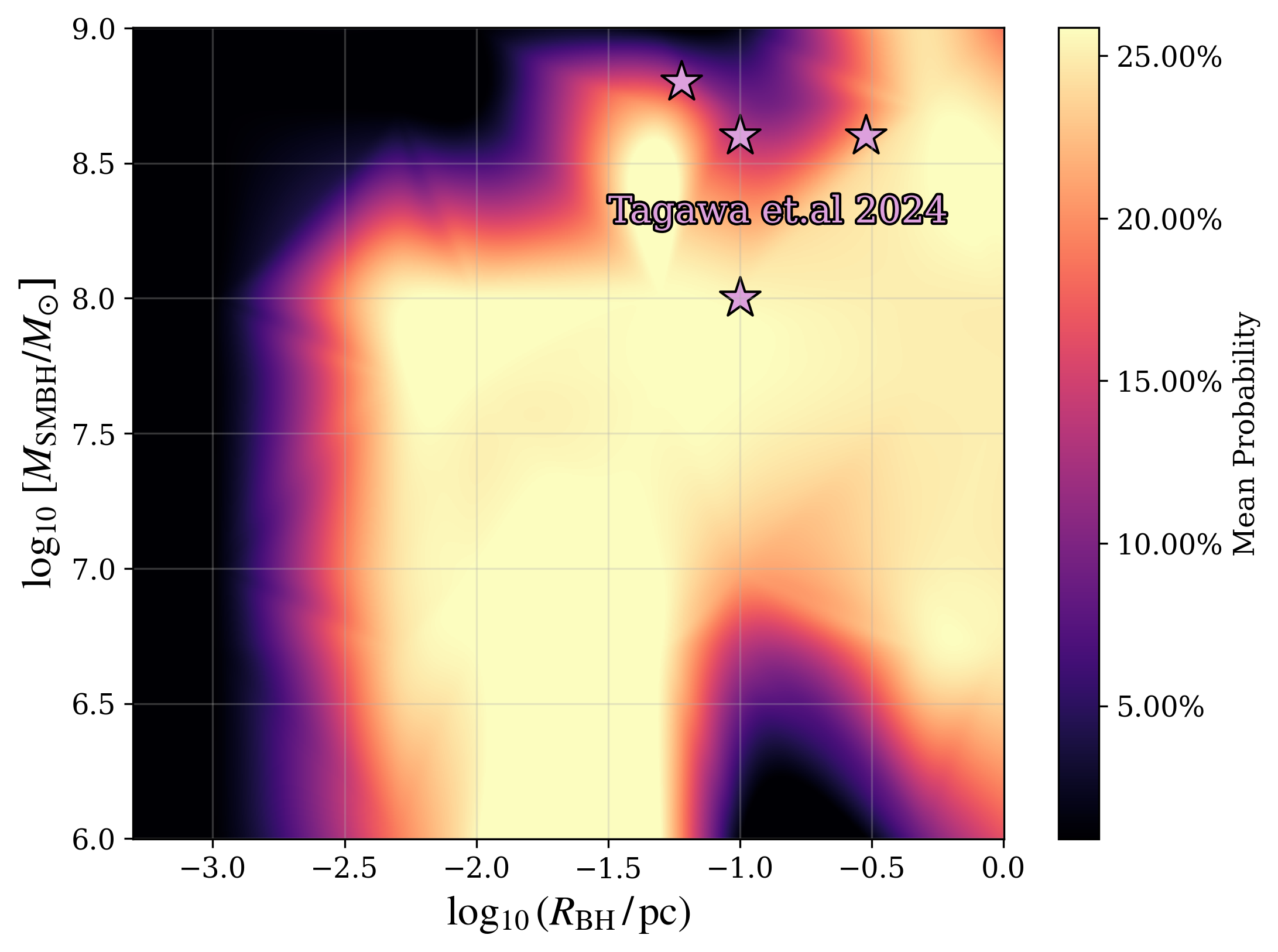}
        \caption{}
        \label{fig:tgw24b}
    \end{subfigure}    
    \caption{Constraints on the presence of a flare assuming the TWG24 model.
    (Left) Mean detection probability across all parameter configurations, illustrating the best-case detection scenario for each $(M_{\mathrm{BH}},\log_{10}(M_{\mathrm{smbh}}/M_{\odot}))$ pair. 
    (Right) Mean detection probability across all parameter configurations, illustrating the best-case detection scenario for each $(M_{\mathrm{BH}},\log_{10}(R_{\mathrm{BH}}/\mathrm{pc}))$ pair. 
    The gravitational wave events shown in the figure correspond to those listed in Table \ref{tab:gwevents}. The star symbols indicate ZTF-detected flares that have been proposed as potential counterparts to GW events \cite{Graham2023}, with their inferred properties of the emission in the shock cooling emission scenario reported by \citet{tagawa_phe_model}.
    Dark regions represent areas of low detection probability, whereas bright regions correspond to regions where flares are highly detectable.
    }
    \label{fig:tgw24ab}
\end{figure*}

\begin{figure}[h]
    \centering
    \includegraphics[width=0.9\linewidth]{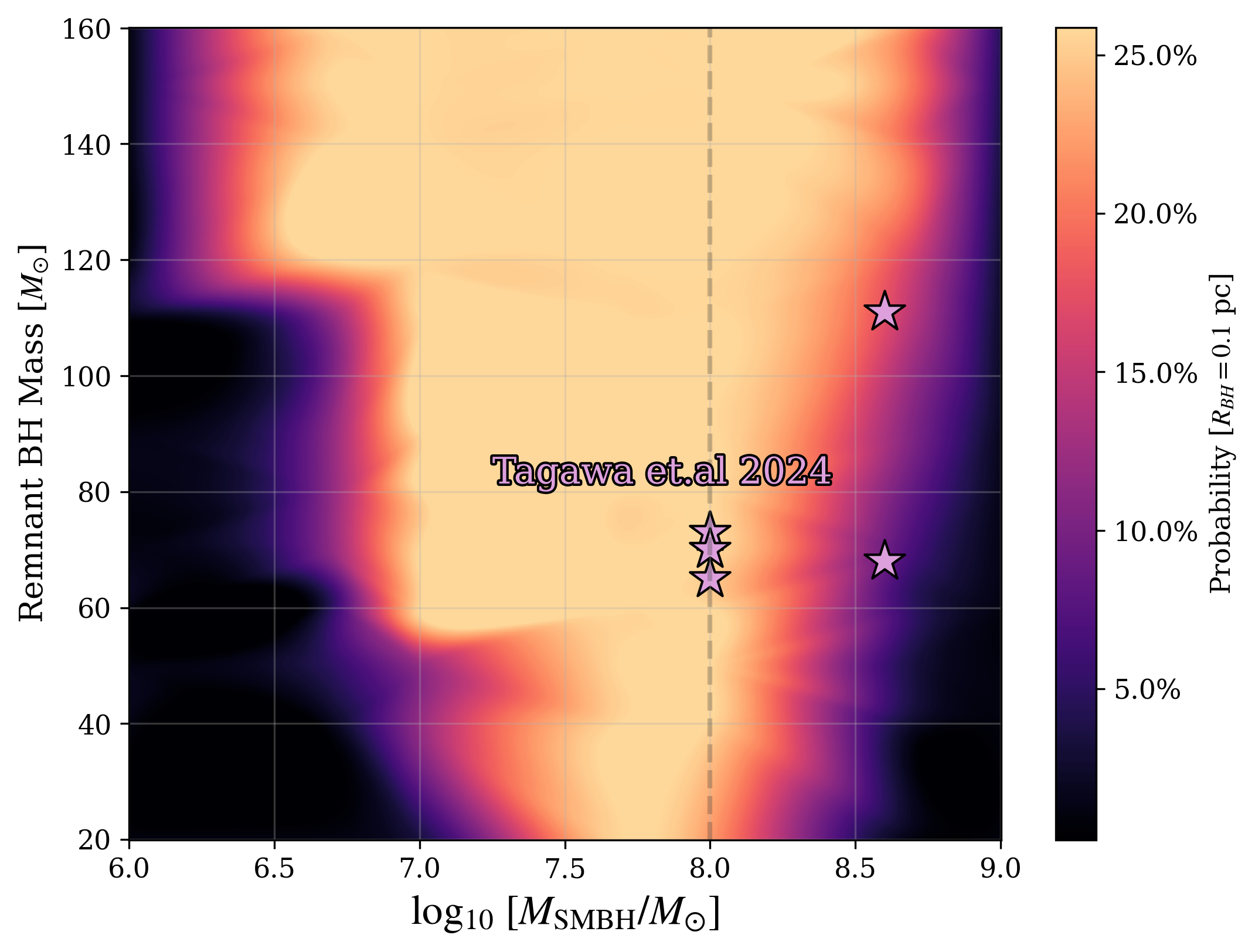}
    \caption{Constraints on the presence of a Flare for the TGW24 model assuming a merget at 0.1 pc. The colorbar shows the estimated likelihood that we would have detected a source for a given remnant Mass and SMBH mass. The star symbols indicate ZTF-detected flares that have been proposed as potential counterparts to GW events \cite{Graham2023}, analyzed under the shock cooling emission scenario at $R_{\mathrm{BH}} = 0.1$ pc as described in \cite{tagawa_phe_model}}
    \label{fig:tgw24c}
\end{figure}

\subsection{The
\textbf{\citeauthor{tagawa_phe_model}(TGW24)} Model
}

In this section, we focus on the constraints on the emission properties predicted by the TGW24 model \citep{tagawa_phe_model}, given our observational strategy.  Figure \ref{fig:tgw24a} reveals a trend broadly consistent with the constraints from the JRR-I model (Figure \ref{fig:jrr1_massb}), showing a slight preference for AGNs hosting SMBHs with masses in the range of  $M_{SMBH} \sim 10^{7} -10^{8} M_{\odot}$.

The mean probability remains below 10\% for all SMBHs with masses greater than $5 \times 10^8\,M_\odot$. Unlike JRR-I, where jet formation is explicitly constrained by AGN disk thickness, the TGW24  scenario assumes jet activity persists after the merger as long as there is sufficient circum-BH gas. Therefore, the AGN mass primarily influences observables such as the duration of the flare and the intrinsic brightness of the host.

Figure \ref{fig:tgw24b} reveals a clear preference for mergers occurring within a radial range of $0.01$–$0.1$ parsec. In the TGW24 model, this region offers a balance of a rapid emission peak and sufficiently long flare durations, making them favorable for early detection. Mergers occurring much closer to the SMBH, at distances below $0.005$ pc, tend to produce very short flares with durations under 10 days, rendering them difficult to detect with our cadence. On the other end, mergers occurring farther out, near $\sim$1 pc, are associated with extended flares, exhibiting a peak happening 40 to 300 days after the merger and durations extending beyond 90 days. 

Our constraints are consistent with the inferred model parameters for the shock cooling emission scenario proposed by \cite{tagawa_phe_model}, based on the candidates reported in \cite{Graham2023}, which are marked as stars in Figure \ref{fig:tgw24b} and \ref{fig:tgw24c}. All the candidates from \cite{tagawa_phe_model} in Figure \ref{fig:tgw24b} fall within the high-probability region ($p_{mean} >5\%$), supporting the idea that the $0.01-0.1$ pc regime provides the most favorable physical conditions for generating observable flares under our survey strategy. This results from the interplay between relatively short delay times and long enough durations for detection.

Figure \ref{fig:tgw24c} provides additional insights into the detectability landscape by showing the dominant contributions to the total probability distribution. Merger configurations at $R_{\mathrm{BH}} = 0.1$ pc consistently exhibit detection probabilities above 25\%. We also find that mergers involving remnant black holes with masses below $60 M_\odot$ are more likely to be detected when occurring around SMBHs in the $5\times 10^{7}-$$5\times10^8 M_\odot$ range. In contrast, more massive remnant BHs are favored around slightly less massive SMBHs in the $10^{6.5}$–$10^8 M_\odot$ range. These results indicate that both radial location and central SMBH mass are key parameters governing the detectability and emission properties of BBH merger counterparts in AGN disks.


\begin{figure*}
    \centering
    \includegraphics[width=1\linewidth]{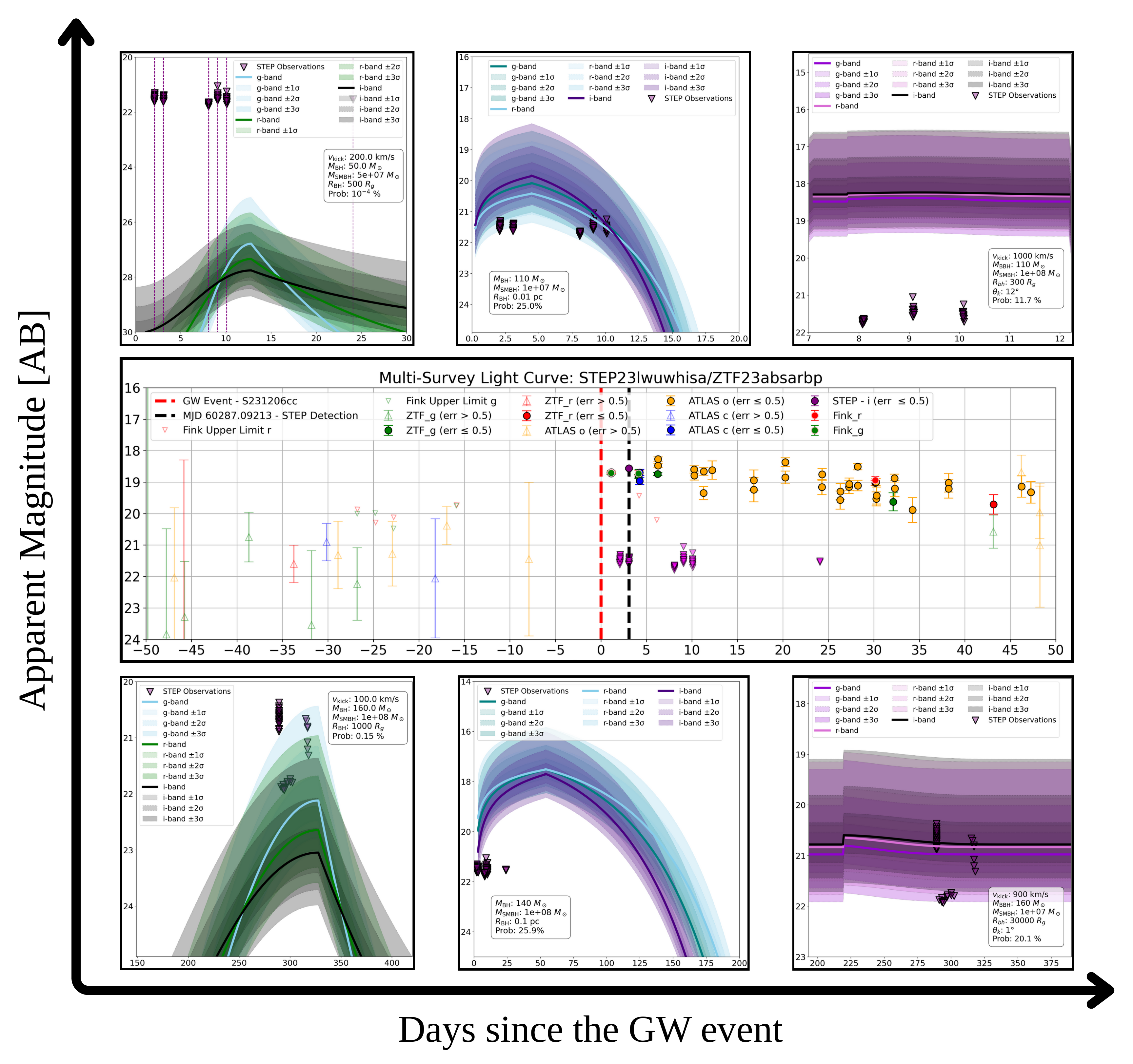}
    \caption{ Observational light curve data from ATLAS, ZTF, Fink, and STEP for the candidate STEPlwuwhisa, initially identified during the search for an electromagnetic counterpart to the GW event S231206cc but later classified as a supernova (ZTF23absarbp). The x-axis shows time since the GW trigger, and the y-axis indicates apparent magnitude (AB). Purple triangles represent the 3-$\sigma$ limiting magnitudes from T80-South follow-up observations across multiple epochs.
    Each panel presents model light curves generated under different parameter configurations for the MCK19 (top and bottom left), TGW24 (top and bottom middle), and JRR-I (top and bottom right) scenarios. The displayed detection probabilities reflect the likelihood of observing a flare under the given configuration. Shaded regions correspond to the model-predicted apparent magnitude ranges, including $\pm1$, $\pm2$, and $\pm3\sigma$ uncertainties due to the luminosity distance measurement.
    Time is referenced relative to the GW trigger, with the bottom row showing early-time flares and the top row showing late-time flares associated with delayed emission.}
    \label{fig:lcall}
\end{figure*}

\subsection{Detectability}
Our investigation highlights the potential for binary black hole mergers occurring within AGN disks to produce observable electromagnetic counterparts, and demonstrates how multi-model constraints can inform optimized follow-up strategies. By exploring three physically motivated models—TGW24, JRR-I, and MCK19, we identify key insights into the detectability of such events.

As expected, we find that detectability is primarily governed by the interplay between the flare’s delay time, its duration and brightness. Our analysis shows that short-delay events often have correspondingly short durations, requiring immediate and high-cadence follow-up observations in the days following the merger. Conversely, flares with longer delay times tend to exhibit prolonged emission, allowing for more temporally sparse but extended monitoring. Most detectable configurations, in both the TGW24 and JRR-I models, predict delay times peaking at $<50$ days, reinforcing the need for extensive early follow-up within the first two months post-merger. In contrast, the MCK19 model yields a maximum detection probability of less than 1\%, primarily due to its intrinsically fainter emission, which falls below typical survey sensitivity limits.

With the upcoming Legacy Survey of Space and Time (LSST; \citealt{verarubin2019ApJ...873..111I}), scheduled to begin operations in late-2025, time-domain astronomy will enter a new era of wide-field, high-cadence optical surveys. While LSST will play a central role in transient discovery, particularly long-duration merger flares, its standard cadence may not be optimal for capturing short-timescale events associated with early-time flare emission ($t_{\mathrm{delay}} < 50$ days). 
Our main strategy with T80-South is designed  to complement LSST by conducting  high-cadence follow-up of transient candidates within the gravitational wave skymap region. This will help close the detection gap for rapidly evolving optical flares that LSST might otherwise miss due to its standard revisit time. In addition, advanced machine learning classifiers trained on archival AGN variability data could also help the discrimination between a merger flare and intrinsic AGN variability at early times \citep{Leoni2022}. This would allow for more efficient prioritization of follow-up targets and better allocation of observational resources. 

The host AGN mass plays a significant role in shaping the emission profile and our ability to detect it. Assuming the BBH event happened on an AGN disk,  if a merger flare occurred but was not detected under our current strategy, it may have originated in a massive AGN host ($M_{\mathrm{SMBH}} \gtrsim 10^9\,M_\odot$), where optical emission is more likely to be obscured or diluted by the bright accretion disk. All models considered in this study favor host SMBH masses below $10^9\,M_\odot$, with TGW24 specifically preferring merger locations that may correspond to migration traps located around $500$–$700\,R_g$ (e.g., \citealt{Bellovary2016,Yang2019}), where embedded BBHs are most likely to coalesce. The JRR-I model exhibits greater variability in detectability as a function of $R_{\mathrm{BH}}$, but still favors mergers in similar radial ranges around AGNs with $M_{\mathrm{SMBH}} \sim 10^7$–$10^8\,M_\odot$. The overlapping regions in parameter space suggest that these moderate-mass AGNs remain the most promising environments for detecting optical counterparts, under the assumption of a fixed accreation rate. 

In the context of the JRR-I sceario, detectability is strongly influenced by the kick velocity of the remnant black hole. Our analysis indicates that larger kick velocities result in shorter delay times, enhancing the chances of detecting prompt flares. In contrast, systems with smaller kicks may produce more luminous emission due to prolonged interaction with the disk, but the associated long delays reduce the likelihood to suceffuly link it as an EM counterpart. In the future, more precise constraints on BH kick velocities—enabled by improved gravitational wave parameter estimation from the final stages of a GW signal \citep{Mahapatra2024}, will allow for more accurate predictions of EM counterpart time-scale.

To date, most follow-up campaigns have relied on a generic “fast and deep” strategy, without incorporating astrophysical priors related to the AGN population or GW-derived parameters. A more definitive strategy remains to be formulated, one that dynamically prioritizes regions based on the AGN distribution and adapts follow-up cadence, filter and depth according to key inferred parameters from GW data (such as remnant mass and kick velocity).  The next gravitational wave observing run (O5), expected to begin in 2027, will provide a critical opportunity to identify electromagnetic counterparts to BBH mergers. Developing such data-informed approach will be critical for enhancing the efficiency of counterpart searches in the upcoming multimessanger era for BBH systems.

During our search for an electromagnetic counterpart to the GW event S231206cc, we identified a promising transient candidate, STEPlwuwhisa, using the STEP pipeline. This source was initially flagged due to its temporal and spatial coincidence with the GW localization region. To investigate its nature, we compiled photometric data from multiple surveys, including ATLAS, ZTF, Fink, and T80-South (STEP). The candidate was independently classified as a likely supernova by the Fink broker's supernova classification module \citep{Leoni2022}. The resulting multi-band light curve is shown in Figure~\ref{fig:lcall}, alongside theoretical merger flare light curves predicted by various BBH emission models. Both early- and late-time flare scenarios are represented, enabling a visual comparison between the observed limiting magnitudes and the detection probabilities calculated by \texttt{Teglon}.

\texttt{Teglon} demonstrates the potential of a versatile, open-source framework to guide the next generation of multimessenger follow-up planning and analysis efforts with model-driven strategies (Dave et al. in prep). We implemented three BBH's flare mechanisms into the software—making it the first analysis tool to integrate multiple BBH emission models in a unified detectability pipeline.  In future work, we plan to expand this framework to include additional models, such as  \cite{juan_novo}, which provides, through the kick velocity, a possible joint constraints on remnant mass and spin. This will enable more comprehensive parameter space coverage and improve predictions for electromagnetic counterparts. To further facilitate BBH follow-up campaigns, \texttt{Teglon} is expected to include the AGN catalog from \cite{Secrest2015}, which contains 1.4 million AGNs selected from the AllWISE catalog \citep{Wright2010wise}. For redshifts $z < 0.1$, this catalog is estimated to be $>$90\% complete, making it a robust resource for prioritizing likely AGN hosts within the LVK localization volume. Incorporating this catalog will allow for an alternative galaxy-weighting scheme focused on AGN environments, thereby increasing the probability of a successful counterpart identification for close events.

\section{Conclusion}\label{sec:Conclusion}

In this work, we conducted an optical follow-up of the BBH merger event S231206cc using the T80-South telescope, focusing on AGN-hosted environments as promising sites for electromagnetic counterparts. Observations were conducted at approximately 2, 3, 8, 9, 10, and 24 days after the GW trigger to capture potential early-time flares. To explore delayed emission scenarios, the same fields were revisited between 289 and 318 days post-merger. Our optical analysis shows that :

\begin{enumerate}
    \item  Given the STEP transient candidate selection criteria and EM counterpart filtering process described in Section 4, we were able to rule out all known transient sources detected within the 90th percentile localization of the S231206cc event. All candidates exhibited prior variability before the GW detection, or were visually classified as artifact. 
    \item We used the 3$\sigma$ limiting magnitudes of our observations to place constraints on the parameter space of three different BBH merger scenarios, based on the predicted flare brightness and timescales from each model. Our analysis indicates that detectable optical flares are most likely to occur when the remnant black hole interacts with the AGN disk at distances of 0.01–0.1 pc from SMBHs in the $10^7$–$10^8\,M_\odot$ range, with delay times shorter than 50 days. We rule out all merger configurations occurring in AGNs with $M_{\mathrm{SMBH}} \gtrsim 10^9\,M_\odot$ under our observational constraints.
    \item For the JRR-I model, detectability strongly depends on the kick velocity of the remnant black hole. Our results show that we are mostly sensitive to larger kicks, which result in shorter delay times and increase the likelihood of detecting prompt flares. This strong dependence highlights the need for tighter constraints on BH kick velocities from future GW data to improve EM follow-up strategies.
    \item Expanding \textit{Teglon} to include a broader suite of physical models and comprehensive AGN catalogs, will enable more effective sky search and target prioritization.  The integration of these elements will allow the community to transform non-detections into meaningful constraints, and detections into breakthroughs in our understanding of black hole astrophysics and jet-driven transients.

\end{enumerate}

\section*{Acknowledgments}
C.D.K. gratefully acknowledges support from the NSF through AST-2432037, the HST Guest Observer Program through HST-SNAP-17070 and HST-GO-17706, and from JWST Archival Research through JWST-AR-6241 and JWST-AR-5441.
CRB acknowledges the financial support from CNPq (316072/2021-4) and from FAPERJ (grants 201.456/2022 and 210.330/2022) and the FINEP contract 01.22.0505.00 (ref. 1891/22). 
The authors made use of Sci-Mind servers machines developed by the CBPF AI LAB team and would like to thank A. Santos, P. Russano, G. Teixeira and M. Portes de Albuquerque for all the support in infrastructure matters.The ZTF forced-photometry service was funded under the Heising-Simons Foundation grant 12540303 (PI: Graham).

\begin{table*}[!h]
    \centering
    \begin{tabular}{|l|l|l|l|l|l|l|}
    \hline
        \textbf{name} & \textbf{mjd} & \textbf{ra} & \textbf{dec} & \textbf{filter} & \textbf{ZTF\_ID} & \textbf{mag} \\ \hline
        STEP23prlurbku & 60287.10422 & 359.97041249999995 & 3.973586111111111 & i & ZTF18ackepif & 19.696 \\ \hline
        STEP23tdfpohqh & 60287.10422 & 359.5889291666666 & 4.106977777777778 & i & None & 19.87 \\ \hline
        STEP23lkwkxcfe & 60287.10422 & 359.31587499999995 & 3.6335166666666665 & i & ZTF20aaiuerf & 19.355 \\ \hline
        STEP23txqomcjc & 60287.10422 & 359.0008833333333 & 4.044375 & i & ZTF22abesnrq & 19.439 \\ \hline
        STEP23bukxneza & 60287.10422 & 358.781 & 4.922016666666667 & i & None & 19.904 \\ \hline
        STEP23ktaqaisl & 60287.09705 & 347.79332916666664 & -12.03048888888889 & i & None & 19.596 \\ \hline
        STEP23ypacijyc & 60287.06794 & 353.8503375 & -2.0012777777777777 & i & None & 17.396 \\ \hline
        STEP23piqcaevs & 60287.09939 & 353.9284375 & 2.91945 & i & None & 19.245 \\ \hline
        STEP23lxzfmmgs & 60287.05833 & 359.41765 & 7.994638888888889 & i & None & 19.493 \\ \hline
        STEP23dnvdagfa & 60287.05833 & 359.39253749999995 & 8.787044444444444 & i & None & 19.848 \\ \hline
        STEP23poheabqf & 60287.05833 & 358.6494041666666 & 8.629150000000001 & i & None & 19.061 \\ \hline
        STEP23sedezvjc & 60287.07024 & 354.1309416666667 & 0.2974277777777778 & i & ZTF21acabxhr & 17.696 \\ \hline
        STEP23veuqjmgi & 60287.05355 & 1.02915 & 8.109805555555555 & i & ZTF21aceocyz & 18.718 \\ \hline
        STEP23nidciaub & 60287.05355 & 0.9756958333333332 & 8.30311388888889 & i & None & 18.665 \\ \hline
        STEP23glmnbsdh & 60287.05355 & 0.9071958333333332 & 8.63543611111111 & i & None & 19.645 \\ \hline
        STEP23kttsruti & 60287.05355 & 0.7148583333333333 & 8.41215 & i & None & 19.953 \\ \hline
        STEP23kjrbvhfc & 60287.05355 & 0.0372749999999999 & 8.29293611111111 & i & ZTF19abnhkml & 19.344 \\ \hline
        STEP23spebcsgt & 60287.07531 & 3.531020833333333 & 10.570858333333334 & i & ZTF18abwgbgw & 19.074 \\ \hline
        STEP23smvsubvh & 60287.06313 & 0.9536708333333332 & 7.478913888888889 & i & None & 18.197 \\ \hline
        STEP23ulhicpwe & 60287.07273 & 352.42712916666665 & -1.5662055555555556 & i & None & 19.389 \\ \hline
        STEP23snogtshs & 60287.07273 & 351.97507499999995 & -1.8236027777777777 & i & None & 19.071 \\ \hline
        STEP23gpojsrji & 60287.07273 & 351.95409583333327 & -0.7314166666666667 & i & None & 19.939 \\ \hline
        STEP23kosdozjo & 60287.07273 & 351.9042166666666 & -0.7121472222222222 & i & None & 19.565 \\ \hline
        STEP23ucnbllcc & 60287.09213 & 355.5839291666666 & 6.032597222222222 & i & None & 20.053 \\ \hline
        STEP23lwuwhisa & 60287.09213 & 355.24086666666665 & 5.129208333333333 & i & ZTF23absarbp & 18.563 \\ \hline
        STEP23ucxumpso & 60287.10177 & 353.76050416666664 & 4.803830555555555 & i & ZTF18ackeycn & 18.647 \\ \hline
        STEP23tevunalh & 60287.10177 & 353.7544583333333 & 4.897627777777777 & i & None & 17.838 \\ \hline
        STEP23rsanssrh & 60287.10177 & 353.470475 & 4.641072222222221 & i & None & 18.959 \\ \hline
        STEP23ynbdfoff & 60287.10177 & 353.4587833333333 & 4.870955555555556 & i & ZTF23abqntxi & 19.21 \\ \hline
        STEP23rlyqcvaa & 60287.10177 & 353.2889625 & 4.552741666666667 & i & None & 18.244 \\ \hline
        STEP23ejrevvha & 60287.05585 & 2.30345 & 10.297613888888888 & i & None & 18.937 \\ \hline
        STEP23ukftivzm & 60287.05585 & 2.0284166666666663 & 9.717866666666668 & i & None & 18.615 \\ \hline
    \end{tabular}
\caption{List of electromagnetic counterparts candidates to the S231206cc event identified by the STEP-GW follow-up campaign. }
\label{tab:candidates}
\end{table*}
\clearpage

\bibliography{step}{}
\bibliographystyle{aasjournal}

\end{document}